\documentclass[12pt,epsf]{article}
\input epsf.tex
\usepackage[dvips]{graphicx}
\usepackage{amsmath}
\usepackage{amsfonts}
\usepackage{latexsym}
\usepackage{amssymb}
\usepackage{feynmf}
\usepackage{multicol}
\begin{document}

\def\a{\alpha}
\def\b{\beta}
\def\c{\chi}
\def\d{\delta}
\def\e{\epsilon}           
\def\f{\phi}               
\def\g{\gamma}
\def\h{\eta}
\def\i{\iota}
\def\j{\psi}
\def\k{\kappa}                    
\def\l{\lambda}
\def\m{\mu}
\def\n{\nu}
\def\o{\omega}
\def\p{\pi}                
\def\q{\theta}                    
\def\r{\rho}                      
\def\s{\sigma}                    
\def\t{\tau}
\def\u{\upsilon}
\def\x{\xi}
\def\z{\zeta}
\def\D{\Delta}
\def\F{\Phi}
\def\G{\Gamma}
\def\J{\Psi}
\def\L{\Lambda}
\def\O{\Omega}
\def\P{\Pi}
\def\Q{\Theta}
\def\S{\Sigma}
\def\U{\Upsilon}
\def\X{\Xi}
\def\del{\partial}


\def\ca{{\cal A}}
\def\cb{{\cal B}}
\def\cc{{\cal C}}
\def\cd{{\cal D}}
\def\ce{{\cal E}}
\def\cf{{\cal F}}
\def\cg{{\cal G}}
\def\ch{{\cal H}}
\def\ci{{\cal I}}
\def\cj{{\cal J}}
\def\ck{{\cal K}}
\def\cl{{\cal L}}
\def\cm{{\cal M}}
\def\cn{{\cal N}}
\def\co{{\cal O}}
\def\cp{{\cal P}}
\def\cq{{\cal Q}}
\def\car{{\cal R}}
\def\cs{{\cal S}}
\def\ct{{\cal T}}
\def\cu{{\cal U}}
\def\cv{{\cal V}}
\def\cw{{\cal W}}
\def\cx{{\cal X}}
\def\cy{{\cal Y}}
\def\cz{{\cal Z}}










\def\bo{{\mathpalette\bop{}}}                        
\def\pa{\partial}                              
\def\de{\nabla}                                       
\def\dell{\bigtriangledown} 
\def\su{\sum}                                         
\def\pr{\prod}                                        
\def\iff{\leftrightarrow}                      
\def\conj{{\hbox{\large *}}} 
\def\lconj{{\hbox{\footnotesize *}}}          
\def\dg{\sp\dagger} 
\def\ddg{\sp\ddagger} 

\def\det{\mathop{\rm det}\nolimits}
\def\slap#1#2{\setbox0=\hbox{$#1{#2}$}
         #2\kern-\wd0{\hbox to\wd0{\hfil$#1{/}$\hfil}}}
\def\sla#1{\mathpalette\slap{#1}}
\def\sp#1{{}^{#1}}                             
\def\sb#1{{}_{#1}}                             
\def\oldsl#1{\rlap/#1}                 
\def\Sl#1{\rlap{\hbox{$\mskip 3 mu /$}}#1}     
\def\SL#1{\rlap{\hbox{$\mskip 4.5 mu /$}}#1}   
\def\Tilde#1{\widetilde{#1}}                   
\def\Hat#1{\widehat{#1}}                       
\def\Bar#1{\overline{#1}}                      
\def\bra#1{\Big\langle #1\Big|}                       
\def\ket#1{\Big| #1\Big\rangle}                       
\def\VEV#1{\Big\langle #1\Big\rangle}                 
\def\abs#1{\Big| #1\Big|}                      
\def\sbra#1{\left\langle #1\right|}            
\def\sket#1{\left| #1\right\rangle}            
\def\sVEV#1{\left\langle #1\right\rangle}      
\def\sabs#1{\left| #1\right|}                  
\def\leftrightarrowfill{$\mathsurround=0pt \mathord\leftarrow \mkern-6mu
        \cleaders\hbox{$\mkern-2mu \mathord- \mkern-2mu$}\hfill
        \mkern-6mu \mathord\rightarrow$}
\def\dvec#1{\vbox{\ialign{##\crcr
        \leftrightarrowfill\crcr\noalign{\kern-1pt\nointerlineskip}
        $\hfil\displaystyle{#1}\hfil$\crcr}}}          
\def\hook#1{{\vrule height#1pt width0.4pt depth0pt}}
\def\leftrighthookfill#1{$\mathsurround=0pt \mathord\hook#1
        \hrulefill\mathord\hook#1$}
\def\underhook#1{\vtop{\ialign{##\crcr                 
        $\hfil\displaystyle{#1}\hfil$\crcr
        \noalign{\kern-1pt\nointerlineskip\vskip2pt}
        \leftrighthookfill5\crcr}}}
\def\smallunderhook#1{\vtop{\ialign{##\crcr      
        $\hfil\scriptstyle{#1}\hfil$\crcr
        \noalign{\kern-1pt\nointerlineskip\vskip2pt}
        \leftrighthookfill3\crcr}}}
\def\claw#1{{\vrule height0pt width0.4pt depth#1 pt}}
\def\leftrightclawfill#1{$\mathsurround=0pt \mathord\claw#1
        \hrulefill\mathord\claw#1$}
\def\overhook#1{\vbox{\ialign{##\crcr\noalign{\kern1pt}
       \leftrightclawfill5\crcr\noalign{\kern1pt\nointerlineskip}
       $\hfil\displaystyle{#1}\hfil$\crcr}}}
\def\der#1{{\pa \over \pa {#1}}}               
\def\fder#1{{\d \over \d {#1}}} 


\def\ha{\frac12}                               
\def\sfrac#1#2{{\vphantom1\smash{\lower.5ex\hbox{\small$#1$}}\over
        \vphantom1\smash{\raise.4ex\hbox{\small$#2$}}}} 
\def\bfrac#1#2{{\vphantom1\smash{\lower.5ex\hbox{$#1$}}\over
        \vphantom1\smash{\raise.3ex\hbox{$#2$}}}}      
\def\afrac#1#2{{\vphantom1\smash{\lower.5ex\hbox{$#1$}}\over#2}}  
\def\dder#1#2{{\pa #1\over\pa #2}}        
\def\secder#1#2#3{{\pa\sp 2 #1\over\pa #2 \pa #3}}          
\def\fdder#1#2{{\d #1\over\d #2}}         
\def\on#1#2{{\buildrel{\mkern2.5mu#1\mkern-2.5mu}\over{#2}}}
\def\On#1#2{\mathop{\null#2}\limits^{\mkern2.5mu#1\mkern-2.5mu}}
\def\under#1#2{\mathop{\null#2}\limits_{#1}}          
\def\bvec#1{\on\leftarrow{#1}}                 
\def\oover#1{\on\circ{#1}}                            
\def\dt#1{\on{\hbox{\LARGE .}}{#1}}                   
\def\dtt#1{\on\bullet{#1}}                      
\def\ddt#1{\on{\hbox{\LARGE .\kern-2pt.}}#1}             
\def\tdt#1{\on{\hbox{\LARGE .\kern-2pt.\kern-2pt.}}#1}   


\def\boxes#1{
        \newcount\num
        \num=1
        \newdimen\downsy
        \downsy=-1.5ex
        \mskip-2.8mu
        \bo
        \loop
        \ifnum\num<#1
        \llap{\raise\num\downsy\hbox{$\bo$}}
        \advance\num by1
        \repeat}
\def\boxup#1#2{\newcount\numup
        \numup=#1
        \advance\numup by-1
        \newdimen\upsy
        \upsy=.75ex
        \mskip2.8mu
        \raise\numup\upsy\hbox{$#2$}}


\newskip\humongous \humongous=0pt plus 1000pt minus 1000pt
\def\caja{\mathsurround=0pt}
\def\eqalign#1{\,\vcenter{\openup2\jot \caja
        \ialign{\strut \hfil$\displaystyle{##}$&$
        \displaystyle{{}##}$\hfil\crcr#1\crcr}}\,}
\newif\ifdtup
\def\panorama{\global\dtuptrue \openup2\jot \caja
        \everycr{\noalign{\ifdtup \global\dtupfalse
        \vskip-\lineskiplimit \vskip\normallineskiplimit
        \else \penalty\interdisplaylinepenalty \fi}}}
\def\li#1{\panorama \tabskip=\humongous                
        \halign to\displaywidth{\hfil$\displaystyle{##}$
        \tabskip=0pt&$\displaystyle{{}##}$\hfil
        \tabskip=\humongous&\llap{$##$}\tabskip=0pt
        \crcr#1\crcr}}
\def\eqalignnotwo#1{\panorama \tabskip=\humongous
        \halign to\displaywidth{\hfil$\displaystyle{##}$
        \tabskip=0pt&$\displaystyle{{}##}$
        \tabskip=0pt&$\displaystyle{{}##}$\hfil
        \tabskip=\humongous&\llap{$##$}\tabskip=0pt
        \crcr#1\crcr}}


\def\phil{@{\extracolsep{\fill}}}
\def\unphil{@{\extracolsep{\tabcolsep}}}


\def\NP{Nucl. Phys. B}
\def\PL{Phys. Lett. }
\def\PR{Phys. Rev. Lett. }
\def\PRD{Phys. Rev. D}
\def\Ref#1{$\sp{#1)}$}


\textheight=8.5in          
\textwidth=6in           



\newcommand\noi{\noindent}
\newcommand\seq{\;\;=\;\;}

\newcommand\ie {{\it i.e.}}
\newcommand\eg {{\it e.g.}}
\newcommand\etc{{\it etc. }}
\newcommand\viz{{\it viz. }}

\newcommand{\bfl}{\begin{flushleft}}
\newcommand{\efl}{\end{flushleft}}
\newcommand{\bc}{\begin{center}}
\newcommand{\ec}{\end{center}}

\def\eq{\begin{equation}}
\def\eqe{\end{equation}}
\def\eqa{\begin{eqnarray}}
\def\eqae{\end{eqnarray}}
\def\be{\begin{equation}}
\def\ee{\end{equation}}
\def\bea{\begin{eqnarray}}
\def\ena{\end{eqnarray}}



\def\nn{\nonumber\\}
\def\up{\uparrow}
\def\da{\downarrow}
\def\imp{~~\Rightarrow~~}
\def\qq{\qquad\qquad\qquad}
\def\ihalf{{\textstyle{i \over 2}}}

\def\({\left(} \def\){\right)} \def\<{\langle } \def\>{\rangle }
\def\[{\left[} \def\]{\right]} \def\lb{\left\{} \def\rb{\right\}}


\def\llsim{\:\lower1.8ex\hbox{$\buildrel<\over{\widetilde{\phantom{
<}}}$}\:}
\def\lsim{\;\raise.4ex\hbox{$\mathop<\limits_{\widetilde{\phantom{<}}}$}\;
}



\newcommand{\ba}{\begin{eqnarray}}
\newcommand{\ea}{\end{eqnarray}}
\def\ni{\noindent}



\font\mybbb=msbm10 at 8pt
\font\mybb=msbm10 at 12pt
\def\bbb#1{\hbox{\mybbb#1}}
\def\bb#1{\hbox{\mybb#1}}
\def\id{\protect{{1 \kern-.28em {\rm l}}}}
\def\I {\bb{1}}
\def\Z {\bb{Z}}
\def\pRe{\bbb{R}}
\def\Re {\bb{R}}
\def\C {\bb{C}}
\def\pC{\bbb{C}}
\def\H {\bb{H}}
\def\de{\nabla}
\def\imp{~~\Rightarrow~~}
\def\st{\star}
\def\dZ2p{\frac{dZ_1}{2\p i}}
\newcommand{\al}{\mbox{$\alpha$}}
\newcommand{\als}{\mbox{$\alpha_{s}$}}
\newcommand{\lm}{\mbox{$\mbox{ln}(1/\alpha)$}}
\newcommand{\bi}[1]{\bibitem{#1}}
\newcommand{\fr}[2]{\frac{#1}{#2}}
\newcommand{\sv}{\mbox{$\vec{\sigma}$}}
\newcommand{\gm}{\mbox{$\gamma_{\mu}$}}
\newcommand{\gn}{\mbox{$\gamma_{\nu}$}}
\newcommand{\Le}{\mbox{$\fr{1+\gamma_5}{2}$}}
\newcommand{\R}{\mbox{$\fr{1-\gamma_5}{2}$}}
\newcommand{\GD}{\mbox{$\tilde{G}$}}
\newcommand{\gf}{\mbox{$\gamma_{5}$}}
\newcommand{\om}{\mbox{$\omega$}}
\newcommand{\Ima}{\mbox{Im}}
\newcommand{\Rea}{\mbox{Re}}
\newcommand{\Tr}{\mbox{Tr}}
\newcommand{\bbeta}[2] {\mbox{$\bar{\beta}_{#1}^{\hspace*{.5em}#2}$}}
\newcommand{\homega}[2]{\mbox{$\hat{\omega}_{#1}^{\hspace*{.5em}#2}$}}
\def\vecnab{\vec{\nabla}}
\def\vx{\vec{x}}
\def\vy{\vec{y}}
\def\arrowk{\stackrel{\rightarrow}{k}}
\def\kbar{k\!\!\!^{-}}
\def\karrow{k\!\!\!{\rightarrow}}
\def\arrowl{\stackrel{\rightarrow}{\ell}}
\def\var{\varphi}
\def\ttZ{\tilde{\tilde{Z}}}
\def\ttz{\tilde{\tilde{z}}}
\def\tz{\tilde{z}}
\def\tZ{\tilde{Z}}
\def\tq{\tilde{\q}}
\def\del{\partial}
\def\stam{\stackrel{\rightarrow}{\m}}
\def\stab{\stackrel{\rightarrow}{\b}}
\def\in{\infty}





\begin{center}
{\Large\bf The infrared $R^\ast$ operation} \\[8mm] 
 S. Larin$^{a}$ and P. van Nieuwenhuizen $^b$ \\ [3mm]
\begin{itemize}
\item[$^a$]
 Institute for Nuclear Research of the
 Russian Academy of Sciences,   \\
 60th October Anniversary Prospect 7a,
 Moscow 117312.\\
E-mail: larin@ms2.inr.ac.ru 
\item[$^b$]
C.N.Yang Institute for 
Theoretical 
Physics, Stony Brook, NY 11790.\\
 E-mail: vannieu@insti.physics.sunysb.edu
\end{itemize}
\end{center}

\begin{abstract}
We describe the infrared R-operation for subtraction of infrared
divergencies in Feynman diagrams.
\end{abstract}

\newpage

Massless particles can introduce infrared divergences (IRD) in
Feynman graphs.  In Minkowski space there are IRD in the $S$
matrix for QED due to the emission of soft real photons, and also
in graphs with virtual photons.  These IRD cancel in the cross
section~\cite{nord}.  Furthermore, in QCD and also in QED in the
approximation that quarks are massless, one has the situation
that massless particles couple to themselves or to other massless
particles, and this leads to further IRD in the $S$ matrix in
Minkowski space, the so-called collinear divergences.  These
cancel also if one averages over the color of the incoming
particles and sums over the momenta of the initial states with
soft collinear particles \cite{kin}, or if one uses factorization
methods.  In this note we discuss IRD in Green's functions in
Euclidean space; they are unrelated to the IRD in Minkowski space
which occur in the $S$ matrix.

Four-dimensional quantum field theories with only dimensionless
coupling constants contain in Euclidean space for generic external
momenta no IRD. Thus in the proof of renormalizability of proper
graphs in QED, all divergences which one encounters (in
Euclidean space) are ultraviolet divergences (UVD).  The $Z$
factors contain thus only information about the small-distance
behavior, and for this reason they can be used to construct
running coupling constants. The same applies to QCD and massive
quarks. Even if one sets the masses of all particles to zero, it
remains true that for generic Euclidean external momenta the
proper Green function in four dimensions are free from
IRD\footnote{In dimensional regularization one sets the following
gluon selfenergy graph \begin{fmffile}{fmfpic17}
\parbox{35mm}{
\begin{fmfgraph*}(12,8)
\fmfleft{i} \fmfright{o} \fmf{vanilla}{i,v,v,o}
\end{fmfgraph*}}
\end{fmffile}\hspace{-2.5cm} to zero. This graph has an UVD but not an IRD, and
its vanishing should be considered as the result of a
computation, not as an independent rule.  (One may for example
replace ${1 \over k^2}$ by ${m^2 \over k^2 (k^2 + m^2)} + {1
\over (k^2 + m^2)}$, and one finds then that the sum of both
integrals vanishes). Similarly
\begin{fmffile}{fmfpp1}
\parbox{20mm}{
\begin{fmfgraph*}(12,8)
  \fmfleft{i} \fmfright{o} \fmf{vanilla}{i,v,o}
  \fmf{vanilla,label=$\bullet$,label.dist=0}{v,v}
\end{fmfgraph*}}
\end{fmffile} \hspace{-0.5cm} $\sim \int d^4k/k^4$ vanishes, but now the IRD (evaluated
at $n>4$) cancels the UVD (evaluated at $n<4$). Again this
cancellation follows from the rules of dimensional
regularization.}.

Spontaneously broken field theories have in general dimensionful
coupling constants.  For example, the
$O (2)$ Goldstone model and the $SU(2)$ Higgs model contain an interaction term $\l v \s^3$
with $\l v$ a dimensionful coupling constant.  Nevertheless these
four-dimensional Goldstone and Higgs models
do not contain any IRD in the Euclidean proper graphs; this is due to
the Goldstone theorem which
states that proper selfenergies $\Pi (p)$ for Goldstone bosons vanish at
$p^2 =0$ even when loops with
massive scalars $\s$ contribute to the Goldstone boson selfenergy.

However for such theories as $\l
\varphi^4 + h \varphi^3$ theory in four dimensions with a
superrenormalizable dimensionful coupling constant $h$, IRD
do occur.  A simple graph which shows this explicitly is as follows
\begin{equation}
\includegraphics
[draft=false, height=1.5in,width=5in,keepaspectratio] {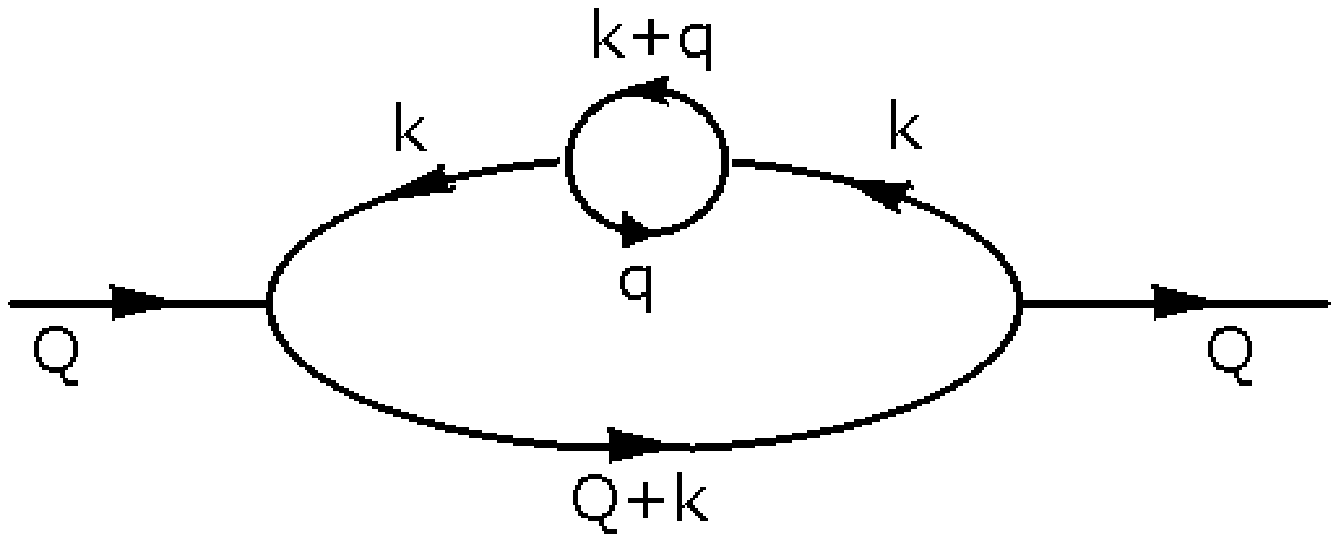}
\end{equation}
We have a massless boson in the larger loop with loop momentum $k$ and a
selfenergy insertion with loop
momentum $q$ due to a massive scalar.  The two massless propagators
$k^{-2}$ lead to a logarithmic IRD $\int d^4 k / k^4$
  because now the proper selfenergy of the massive scalar $\Pi (k) = \int
{1 \over q^2 +
m^2} {1 \over  (k-q)^2 + m^2} d^4 q$ does not vanish for small $k$.

Also at exceptional momenta IRD in Euclidean space can occur as the
following example shows
\begin{equation}
\includegraphics
[draft=false, height=2in,width=5in,keepaspectratio] {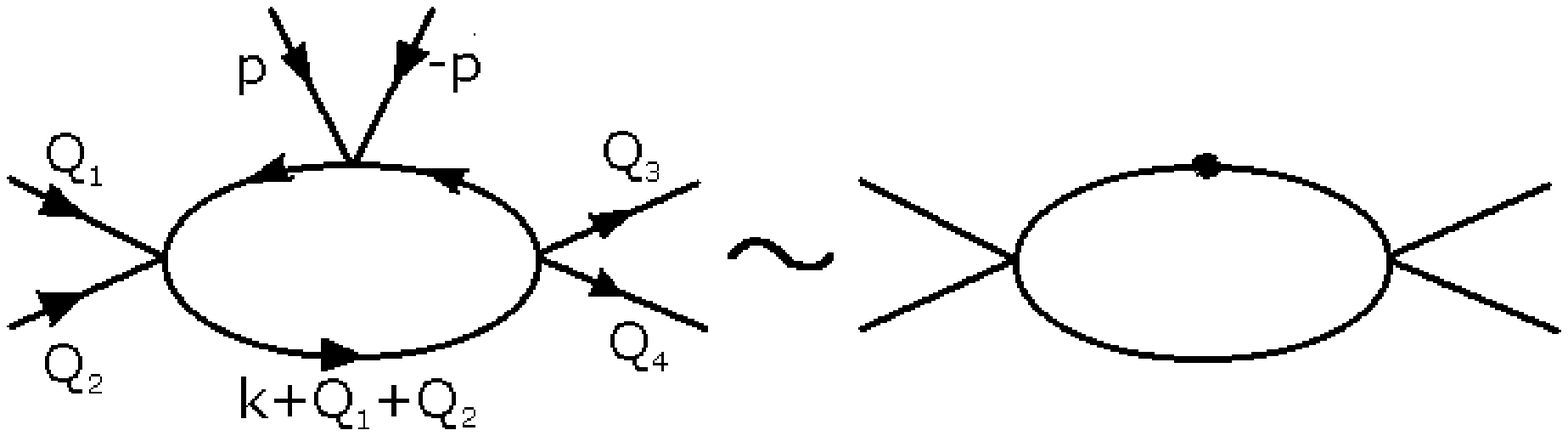}
\end{equation}
The two external lines in the middle carry momenta $p$ and $-p$ and one
can also write them as a $\varphi^2$
-insertion.  The loop integral yields $\int d^4 k {1 \over k^4} {1 \over
(k + Q)^2}$ with $Q=Q_1 + Q_2$, and contains clearly an
IRD at $k=0$.

Another area where IRD create problems is in the calculations of
higher-loop $\beta$
functions.  Suppose we want to compute the divergences in the following
4-point graph
\begin{equation}
\includegraphics
[draft=false, height=1.5in,width=5in,keepaspectratio] {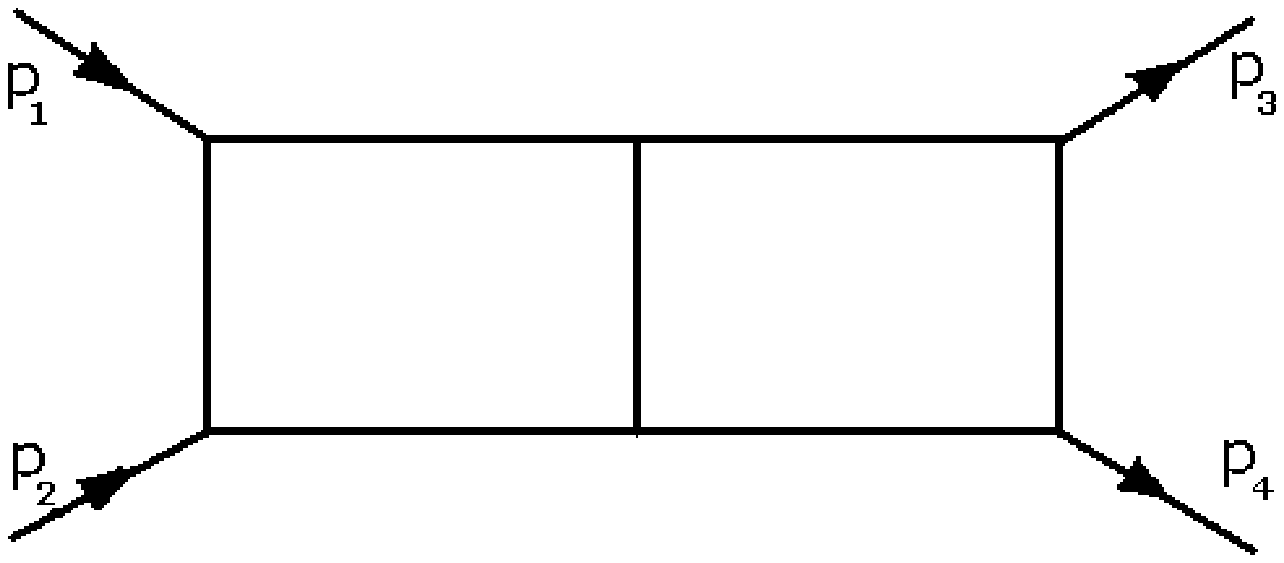}
\end{equation}
It simplifies the calculation a great deal if one sets $p_3 = p_4
=0$. When we set some external momenta to zero, we shall call
this operation ``nullification of momenta".  Nullification of some
of the external momenta makes loop calculations a lot easier but
it creates spurious IRD.  (By spurious we mean that for generic
momenta there are no such infrared divergences).  To compute the
UVD in this graph after having put $p_3 = p_4 =0$, one must first
remove the spurious IRD~\cite{Vlad}.

The conclusion is that in four-dimensional Euclidean space Feynman
graphs may contain IRD
in the following cases\\
(i) if there are superrenormalizable coupling constants such as $\l
\varphi^3$ interactions in $D=4$.\\
(ii) at special external momenta.  More precisely, when the sum of some
external momenta is equal to the
sum of some (or none) of the other external momenta; in particular when
some external momenta are nullified.   (In
the example with external momenta $+p$ and $-p$, the sum of these two
external momenta vanishes).

When there are both massless and massive particles in the theory,
the situation is much more complicated, and one must proceed by
studying each theory separately. For example the interaction
$g\varphi\chi^2$ in $D=6$ with a massless field $\varphi$ and a
massive field $\chi$ leads to IRD in the $\varphi$ selfenergy
with two or more closed $\chi$ loops,but if one first
renormalizes and imposes the renormalization condition that
the $\varphi$ self-energy vanishes like $k^2$, the IRD
disappears. This example shows that in general {\bf one should
first renormalize and then study IRD}.

In higher spacetime dimensions the degree of IRD in general
decreases because of the measure $d^D k$, but in lower spacetime
dimensions one encounters more IRD.  In particular, in $D=2$ the
nonlinear sigma models with action $g_{ij} \del_\mu \varphi^i
\del^\mu \varphi^j$ have dimensionless coupling constants (for
example $g_{ij} = \delta_{ij} (1+ g \varphi^2)$ has the
dimensionless coupling constant $g$) but tadpoles and
selfenergies are IRD.  On the other hand, $\l \varphi^3$ theory
in $D=6$ dimensions is renormalizable but not superrenormalizable
(because $\l$ is now dimensionless) and has no IRD.\footnote{As
an example, consider in 6 dimensions a selfenergy graph with a
massless scalar in the loop, and insert into this loop a string
of $M$ selfenergies with massive scalars in the loops.   Then the
propagators yield a factor $({1 \over k^2})^{M+1}$ and the
measure yields $\int d^6 k$, but now each massive renormalized selfenergy
yields a factor $\int d^6 q {1 \over (q^2 +m^2)(k-q)^2 +m^2} \sim
k^2$, and there is indeed no infrared divergence.}

To explain these various results on IRD one can use some simple IR
power counting.  Consider a proper graph with nonvanishing
external momenta $p_i$ in $D$ dimensions.  We shall assume that
the external momenta are nonexceptional by which we mean that
there does not exist a relation \eqa \sum^{M}_{j=1} p_j =
\sum^{N}_{k=1} p_k \eqae for any $M \geq 0$ and $N>0$ other than
overall energy-momentum conservation.  The propagators contain
loop momenta and external momenta.  Choosing a particular
momentum flow through the diagram, there are ``soft propagators"
with only loop momenta and ``hard propagators" with a combination
of loop momenta and external momenta.  (There are no propagators
with only a combination of external momenta since the graph is
proper).  For vanishing loop momenta, the hard propagators do not
become singular if we do not have exceptional external momenta.
Hence, for nonexceptional values, the external momenta provide an
infrared cut-off for the hard propagators.   We can then determine
the IRD which occur if one or more loop momenta tend to zero by
shrinking all hard propagators to a point.  The following example
in $\l \varphi^4$ theory illustrates this procedure
\begin{equation}
\label{one1115}
\includegraphics
[draft=false, height=1.5in,width=5in,keepaspectratio] {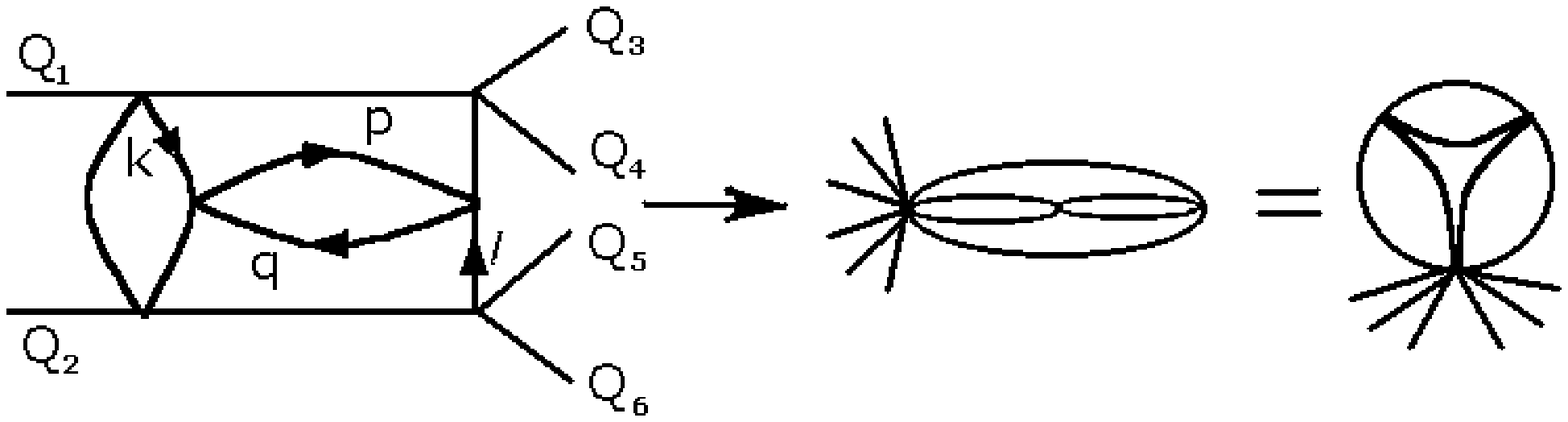}
\end{equation} The contracted graphs are still proper when the
original graph was proper.  Because the hard propagators form a
connected graph, there is only one contracted vertex for a given
proper graph when the external momenta are nonexceptional.

Just as one can count the degree of UVD of a proper graph by UV
counting rules one can also develop IR counting rules.  To
perform $IR$ counting, consider a contracted proper graph with $N$
external lines, $L$ loops, $I$ internal propagators and vertices
$V_j$ with $j$ lines, in addition to the contracted vertex.  Let
$N_i$ be the number of soft lines at the contracted vertex (the
subscript $i$ stands for internal). Because the contracted graph
is still proper the number of soft lines connecting it to the rest
of the graph is at least 2, hence $N_i \geq 2$.  Then the usual
counting rule for the number of loops and the relation which
states that any internal (external) line ends at two (one)
vertices, lead to \eqa && L = I - \left( \sum_j V_j + 1 \right) +
1 \nn && N+ 2I = \sum_j j V_j +(N + N_i) \eqae (In the example
$L=4, N=6, V_4 = 2, I=6$ and $N_i = 4$. Then $L=4=6-2$ and $N+ 2
I = 18 =8 +6+4$). If there are no superrenormalizable couplings,
a vertex $V_j$ in $D$ dimensions carries $D- {1 \over 2} j (D-2)$
momenta attached to it.  (For example, in gauge theory in $D=4$,
the $AA \del A$ vertex carries one momentum and the $AAAA$ vertex
carries no momentum).  When all loop momenta tend to zero, minus
the overall degree of IRD of a contracted graph is given by \eqa
\o_{IR} &=& DL - 2I + \sum V_j \left( D- {1 \over 2} j (D-2)
\right) \nn &=& {1 \over 2} (D-2) N_i \geq D-2 \label{one1117}
\eqae

Therefore in $D \geq 3$ there are in general no overall IRD, but
in $D=2$ all contracted graphs are logarithmically IRD.  In
particular the widely used WZWN models contain IRD.  On the other
hand, gauge theories in 4 dimensions have no IRD, as we already
discussed.

What happens if only some of the loop momenta tend to zero, but
others do not?  If a number $I_H \leq I$ of the internal momenta
are kept hard, one can further contract the original diagram such
that only the soft lines ($I -I_H$ in number) remain.   For
example, one could make the loop momentum of the soft loop at the
top in figure (\ref{one1115}) hard. If one were to make the two
propagators in the loop on top hard, contraction of this new hard
lines would yield a second contracted vertex, but contraction of
one of the loops on the side would still leave only one
contracted vertex.

Let the new hard lines form a subgraph with $L_H$ loops, $I_H$
propagators and $V_{jH}$ vertices.  The remaining degree of IRD
is in this case \eqa && \o_{IR} \; {\rm (subcase)} \; = \o_{IR} -
\Delta \o_{IR} \nn && \Delta \o_{IR} = D L_H - 2I_H + \sum j
V_{jH} \left( D- {1 \over 2} j (D-2) \right) \eqae The subgraph
can be worse IR divergent than the original graph because
$\o_{IR}$ (subcase) is equal to $\o_{IR}$ of the original graph
minus the degree of infrared divergence $\Delta \o_{IR}$ of the
subgraph.  The original set of hard lines together with the new
set of hard lines form a new (possibly disconnected) set of hard
lines, and we can again apply the IR counting rules to the new
contracted graph. In the example in (\ref{one1115}) with $I=6$
soft lines we can make the two soft lines on the right hard. Then
the soft part subdivides as follows
\eqa
\begin{array}{ll} \o_{IR} \; {\rm (subcase)} \;  = 3D-8+(-D+4)=2D-4 \\
\Delta \o_{IR} = D-4+(D-2(D-2))=0
\end{array}
\eqae
The subcase
remains indeed IR finite in $D\geq 3$. As another example, one
may make the loop on top of figure~(\ref{one1115}) hard. Than one
finds that $\Delta\omega_{IR}=D-4+(-D+4)=-D+4$, and
$\omega_{IR}(\mbox{subcase})=3D-8$, which is again IR for $D\geq
3$. However, an example of a subgraph which is divergent while
the minimal graphis not is the following self energy in D=6:
\begin{equation}
\label{flup}
\includegraphics
[draft=false, height=1.5in,width=5in,keepaspectratio] {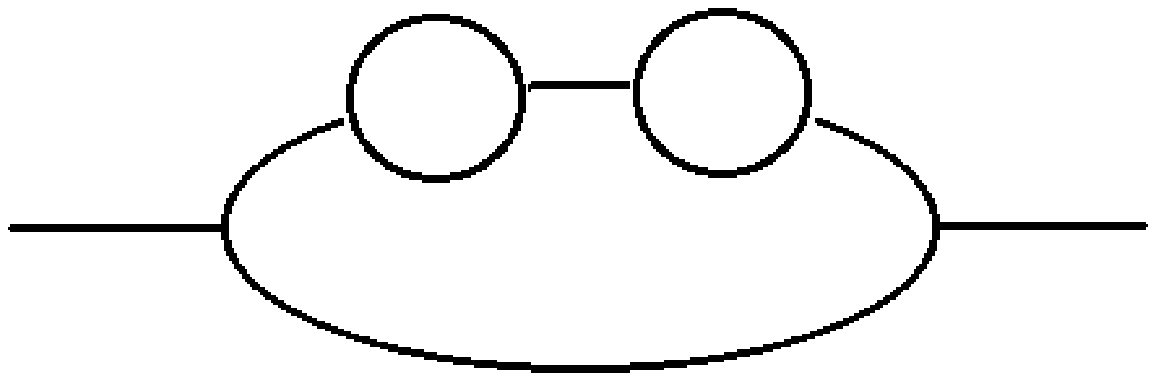}
\end{equation}
The graph is overall IR finite (as we have proven generally) but
contracting the two subloops, one finds an IRD proportional to
$\int d^6k/k^6$. We must thus learn how to subtract IR
subdivergences.

There exists a general scheme for subtracting IRD, which is the
counter part of the so-called $R$-scheme of BPHZ for subtracting
UVD~\cite{bog}. The scheme which subtracts both UVD and IRD is
called the $R^\ast$ scheme~\cite{chet}, and one can formally write
\eqa \label{pochemu} R^\ast = R_{UV} R_{IR} \eqae According to
this prescription, one first removes all IRD of given graph, and
only afterwards subtracts all UVD. This is correct for most
cases, but there are graphs where $R_{UV}R_{IR}$ is not equal to
$R_{IR}R_{UV}$. The question then arises which order is the
correct one, and the answer is that the correct order is
$R_{IR}R_{UV}$, and not~(\ref{pochemu}). We already argued before
that in the case of theories with massive and massless particles
one should first renormalize before extracting IRD. In most of
our examples we shall follow the prescription in~(\ref{pochemu})
because this is technically easier, but we shall also discuss a
5-loop graph where~(\ref{pochemu}) is not correct.

The UVD can be canceled by the usual UV counter terms, but the
IRD are discarded by hand. For the calculation of $\beta$
functions this is no problem because the IR divergences due to
nullification of external momenta were anyhow spurious, but for
field theories such as $h \varphi^3$ in $D=4$ discarding genuine
IRD by hand seems a dubious procedure. One would prefer to have
also an IR renormalization procedure similar to the UV
renormalization procedure, but it does not seem to exist.
Speculations have been made that the sum of infrared divergences
vanishes in the two-dimensional WZWN model (when properly
summed).\cite{bog}  If this does not happen in this model, or in
massless superrenormalizable theories, one would have to exclude
such theories.

In the $R$ subtraction scheme of BPHZ, graphs are expanded into a
Taylor series in the external momenta.  We shall instead use
dimensional regularization to compute both the IRD and the UVD.
Before going on, we make a comment on the relation $\int d^4
k/k^4 =0$ in dimensional regularization.  The reason this
integral vanishes is that it contains both an UVD and an IRD,
whose sum cancels. (One may separately compute the UVD from $\int
d^4 k/(k^2 + m^2)^2$ and the IRD from $\int d^4 k {1 \over k^4}
{1 \over k^2 +m^2}$ and show that their sum cancels).  If one
were to use $\int d^4 k/k^4 =0$ in higher-loop $\beta$ function
calculations, one would drop some UVD, and hence one would make
an error.  In fact, one never encounters the need for setting
$\int d^4 k/k^4 =0$ in the computation of the $\beta$ function at
the one-loop and two-loop level, but at higher loops care is
required not to discard UVD.  The claim is that the $R^\ast$
scheme does not loose UVD even though it sets tadpoles to zero
according to the rules of dimensional regularization.

Let us now explain the infrared subtraction procedure by some
simple examples.  Consider massless $\l \varphi^4$ theory in
$D=4$.  The proper graph
\begin{equation}
\includegraphics
[draft=false, height=1in,width=5in,keepaspectratio] {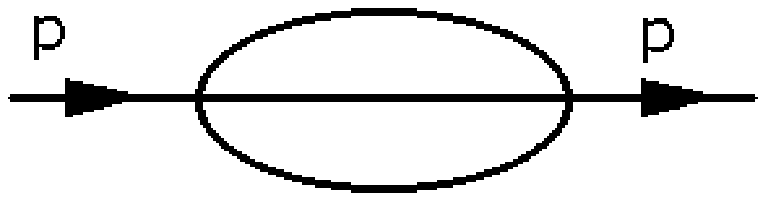}
\end{equation}
contains no IRD at generic $p$, hence there is nothing to
subtract. This is clear after contracting the graph
\begin{equation}
\includegraphics
[draft=false, height=1in,width=5in,keepaspectratio] {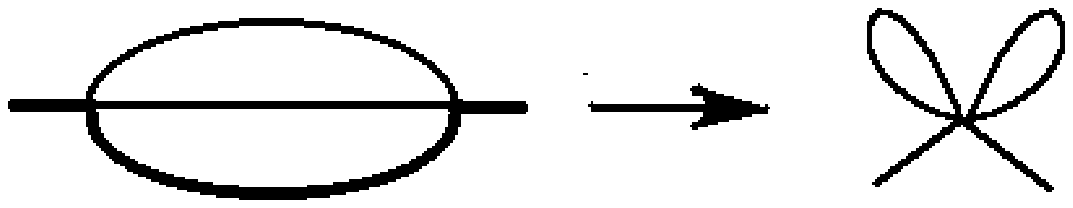}
\end{equation}
Let us introduce an operator
$R_{IR}$ which projects out the infrared finite part from a given
graph.  Then in this example we obtain \eqa R_{IR}\;\;
\begin{fmffile}{fmfq11}\parbox{20mm}{
\begin{fmfgraph*}(12,8)
 \fmfleft{i}
 \fmfright{o}
 \fmf{vanilla,tension=8}{i,v1,v2,o}
 \fmf{vanilla,right}{v1,v2}
 \fmf{vanilla,left}{v1,v2}
\end{fmfgraph*}}
\end{fmffile} \hspace{-0.4cm}= \quad
\begin{fmffile}{fmfq12}\parbox{20mm}{
\begin{fmfgraph*}(12,8)
 \fmfleft{i}
 \fmfright{o}
 \fmf{vanilla,tension=8}{i,v1,v2,o}
 \fmf{vanilla,right}{v1,v2}
 \fmf{vanilla,left}{v1,v2}
\end{fmfgraph*}}\hspace{-0.4cm} =  \quad \mbox{IR-finite}
\end{fmffile}
\eqae

However, consider in $D=4$ the graph 

\begin{equation}
\label{zabil}
\begin{array}{c}\hspace{-3cm}
\begin{fmffile}{fmfq14}\parbox{20mm}{
\begin{fmfgraph*}(50,30)
 \fmfleft{i}
 \fmfright{o}
 \fmf{vanilla,tension=8}{i,v1,v2,o}
 \fmf{vanilla,right,label=$\bullet$,label.dist=0}{v1,v2}
 \fmf{vanilla,left}{v1,v2}
\end{fmfgraph*}}
\end{fmffile}
\end{array}
\end{equation} 
where the dot indicates that two
external momenta have been nullified. There is then clearly a
logarithmic divergence $\int d^4 k/k^4$.  This IRD is due to the
double propagator \begin{fmffile}{fmfq1}\parbox{20mm}{
\begin{fmfgraph*}(12,8)
 \fmfleft{i}
 \fmfright{o}
 \fmf{vanilla}{i,v,o}
 \fmfdot{i}
\fmfdot{v} \fmfdot{o}
\end{fmfgraph*}}
\end{fmffile}\hspace{-0.7cm} and we
can determine the divergence it produces by inserting it into the
simplest graph where it yields an IRD.  So we work in two steps:
we determine first the IR divergent part associated with
\begin{fmffile}{fmfq1}\parbox{20mm}{
\begin{fmfgraph*}(12,8)
 \fmfleft{i}
 \fmfright{o}
 \fmf{vanilla}{i,v,o}
 \fmfdot{i}
\fmfdot{v} \fmfdot{o}
\end{fmfgraph*}}
\end{fmffile}\hspace{-0.7cm} and then we use it to
subtract the IRD from the original graph.  We indicate the IRD
induced by the double propagator
\begin{fmffile}{fmfq1}\parbox{20mm}{
\begin{fmfgraph*}(12,8)
 \fmfleft{i}
 \fmfright{o}
 \fmf{vanilla}{i,v,o}
 \fmfdot{i}
\fmfdot{v} \fmfdot{o}
\end{fmfgraph*}}
\end{fmffile}\hspace{-0.7cm} by $\left(
\hspace{-0.3cm}\begin{fmffile}{fmfq2}\parbox{20mm}{
\begin{fmfgraph*}(12,8)
 \fmftop{i}
 \fmfbottom{o}
 \fmf{vanilla}{i,v,o}
 \fmfdot{i}
\fmfdot{v} \fmfdot{o}
\end{fmfgraph*}}
\end{fmffile}\hspace{-1.1cm}\right)_{IR}$.  Then $\left(
\hspace{-0.3cm}\begin{fmffile}{fmfq2}\parbox{20mm}{
\begin{fmfgraph*}(12,8)
 \fmftop{i}
 \fmfbottom{o}
 \fmf{vanilla}{i,v,o}
 \fmfdot{i}
\fmfdot{v} \fmfdot{o}
\end{fmfgraph*}}
\end{fmffile}\hspace{-1.1cm}\right)_{IR}$ should make the simplest graph with an
\begin{fmffile}{fmfq1}\parbox{20mm}{
\begin{fmfgraph*}(12,8)
 \fmfleft{i}
 \fmfright{o}
 \fmf{vanilla}{i,v,o}
 \fmfdot{i}
\fmfdot{v} \fmfdot{o}
\end{fmfgraph*}}
\end{fmffile}\hspace{-0.7cm} insertion finite as far
as IR divergences go
\begin{equation}
\includegraphics
[draft=false, height=1in,width=5in,keepaspectratio] {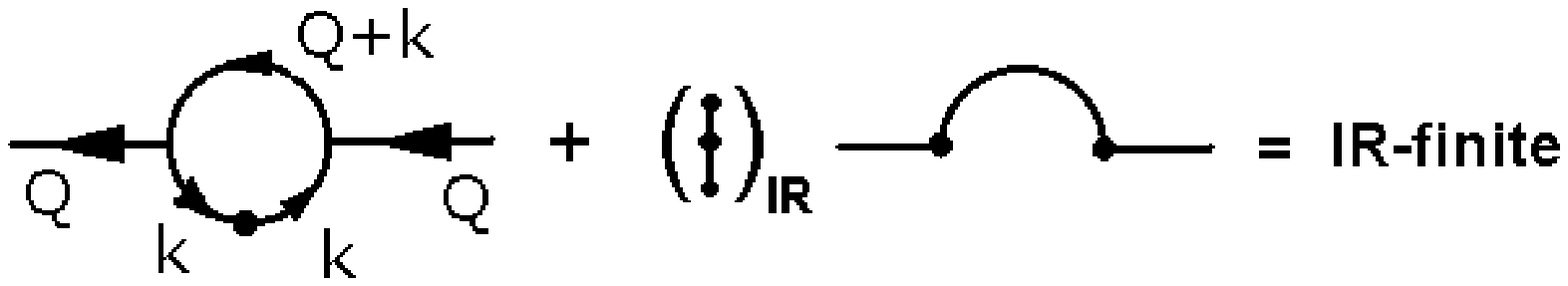}
\end{equation}
Note that we obtain the remainder of the graph after extracting
the IRD by deleting the subgraph which yields the IRD, not by
contracting it. Analytically the meaning of this graphic equation
is as follows \eqa \int d^D k {1 \over k^4 (Q -k)^2} + \left( c_1
{1 \over \e} \right) {1 \over Q^2} =  \;  \mbox{IR-finite} \eqae

We can also formulate the subtraction procedure by replacing ${1 \over
k^4}$ in the original graph by a term with $\delta^4 (k)$
\eqa
{1 \over k^4} \rightarrow {1 \over k^4} + c_1 {1 \over \e} \delta^D (k)
\mu^\e \; , \; \e = D - 4
\eqae
The factor $\mu^\e$ is needed in dimensional regularization to keep the
dimension of the last term the same as that of $k^{-4}$.
Given the rules of dimensional regularization, one can compute $c_1$.

We now return to the original graph in~(\ref{zabil}). As before,
the infrared finite part of the original graph is projected out by
the operator $R_{IR}$, and pictorially one has the following
relation \eqa R_{IR}\left(
\begin{fmffile}{fmfq31}\parbox{20mm}{
\begin{fmfgraph*}(12,12)
 \fmfleft{i}
 \fmfright{o}
\fmf{vanilla,tension=15}{i,v1,v2,o}
 \fmf{vanilla,right,tension=0.3,label=$\bullet$,label.dist=0}{v1,v2}
 \fmf{vanilla,left,tension=0.3}{v1,v2}
\end{fmfgraph*}}
\end{fmffile}\hspace{-0.6cm}
\right) =\;\;
\begin{fmffile}{fmfq29}\parbox{20mm}{
\begin{fmfgraph*}(12,12)
 \fmfleft{i}
 \fmfright{o}
 \fmf{vanilla,tension=15}{i,v1,v2,o}
 \fmf{vanilla,right,tension=0.3,label=$\bullet$,label.dist=0}{v1,v2}
 \fmf{vanilla,left,tension=0.3}{v1,v2}
\end{fmfgraph*}}
\end{fmffile} \hspace{-0.4cm}+\;\;
\left( \hspace{-0.3cm}\begin{fmffile}{fmfq2}\parbox{20mm}{
\begin{fmfgraph*}(12,8)
 \fmftop{i}
 \fmfbottom{o}
 \fmf{vanilla}{i,v,o}
 \fmfdot{i}
\fmfdot{v} \fmfdot{o}
\end{fmfgraph*}}
\end{fmffile}\hspace{-1.1cm}\right)_{IR}\quad
\begin{fmffile}{fmfq30}\parbox{20mm}{
\begin{fmfgraph*}(12,12)
 \fmfleft{i}
 \fmfright{o}
 \fmf{vanilla}{i,v1}
 \fmf{vanilla}{v2,o}
 \fmf{vanilla,left,tension=.3}{v1,v2,v1}
\end{fmfgraph*}} \end{fmffile}\hspace{-0.4cm} =  \quad \mbox{IR-finite} \label{one118} \eqae
Note that one always begins
with the original graph on the right-hand side, and then one adds
terms which subtract the IRD of the various subgraphs of the
original graph.  By convention one always writes plus signs on the
right-hand side.

The UV subtraction scheme can be formulated in the same way.
Consider for example the graph in $D=4$ \begin{equation}
\begin{fmffile}{fmfq44}\parbox{20mm}{
\begin{fmfgraph*}(25,15)
\fmfleft{i} \fmfright{o} \fmftop{t} \fmfbottom{b}
\fmf{vanilla,tension=8}{i,v1} \fmf{vanilla,tension=8}{v2,o}
\fmf{vanilla}{b,v1,t,v2,b} \fmf{vanilla,left=0.4,tension=1}{t,b}
\fmf{vanilla,right=0.4,tension=0}{t,b}
\end{fmfgraph*}}
\end{fmffile}
\end{equation} It has a logarithmic UVD due to the loop in the
middle, and also two logarithmic UV divergences due to the two
2-loop subgraphs on the left and on the right.  (These two
subgraphs are overlapping but that does not modify the
subtraction rules).  Let us introduce an operator $R_{UV}$ which
projects out the UV finite part of a graph.  One obtains then
\begin{equation} R_{UV}\;\;
\begin{fmffile}{fmfq45}\parbox{20mm}{
\begin{fmfgraph*}(10,6)
\fmfleft{i} \fmfright{o} \fmftop{t} \fmfbottom{b}
\fmf{vanilla,tension=8}{i,v1} \fmf{vanilla,tension=8}{v2,o}
\fmf{vanilla}{b,v1,t,v2,b} \fmf{vanilla,left=0.4,tension=1}{t,b}
\fmf{vanilla,right=0.4,tension=0}{t,b}
\end{fmfgraph*}}
\end{fmffile}\hspace{-1cm}\;\;=\;\;
\begin{fmffile}{fmfq46}\parbox{20mm}{
\begin{fmfgraph*}(10,6)
\fmfleft{i} \fmfright{o} \fmftop{t} \fmfbottom{b}
\fmf{vanilla,tension=8}{i,v1} \fmf{vanilla,tension=8}{v2,o}
\fmf{vanilla}{b,v1,t,v2,b} \fmf{vanilla,left=0.4,tension=1}{t,b}
\fmf{vanilla,right=0.4,tension=0}{t,b}
\end{fmfgraph*}}
\end{fmffile}\hspace{-1cm}\;+\; \left(\hspace{-0.3cm}
\begin{fmffile}{fmfq47}\parbox{20mm}{
\begin{fmfgraph*}(10,6)
\fmftop{t} \fmfbottom{b} \fmfdot{t} \fmfdot{b}
 \fmf{vanilla,left=0.4,tension=1}{t,b}
\fmf{vanilla,right=0.4,tension=1}{t,b}
\end{fmfgraph*}}
\end{fmffile} \hspace{-1.3cm}\right)_{UV}
\begin{fmffile}{fmfq48}\parbox{20mm}{
\begin{fmfgraph*}(12,12)
 \fmfleft{i}
 \fmfright{o}
 \fmf{vanilla}{i,v1}
 \fmf{vanilla}{v3,o}
 \fmf{vanilla,left,tension=.3}{v1,v2,v1}
\fmf{vanilla,left,tension=.3}{v2,v3,v2}
\end{fmfgraph*}} \end{fmffile}
\hspace{-1cm}\;\;\;+\; 2\left(\hspace{-0.3cm}
\begin{fmffile}{fmfq49}\parbox{20mm}{
\begin{fmfgraph*}(10,6)
\fmftop{t} \fmfbottom{b} \fmfright{o} \fmfdot{o} \fmfdot{t}
\fmfdot{b} \fmf{vanilla}{t,o,b}
 \fmf{vanilla,left=0.4,tension=1}{t,b}
\fmf{vanilla,right=0.4,tension=1}{t,b}
\end{fmfgraph*}}
\end{fmffile} \hspace{-0.9cm}\right)_{UV}
\begin{fmffile}{fmfq50}\parbox{20mm}{
\begin{fmfgraph*}(12,12)
 \fmfleft{i}
 \fmfright{o}
 \fmf{vanilla}{i,v1}
 \fmf{vanilla}{v2,o}
 \fmf{vanilla,left,tension=.3}{v1,v2,v1}
\end{fmfgraph*}} \end{fmffile}
\end{equation} Again by convention we always use plus signs for the terms
to be subtracted.

After the IRD have been subtracted,
one may subtract the UVD.  One does this for each term on the right-hand
side of the $R_{IR}$ equation
separately.  Consider for example (\ref{one118}).  The subtraction of
UVD proceeds as follows
\begin{equation}
\begin{array}{l}
R_{UV}\left(
\begin{fmffile}{fmfq29}\parbox{20mm}{
\begin{fmfgraph*}(12,12)
 \fmfleft{i}
 \fmfright{o}
 \fmf{vanilla,tension=15}{i,v1,v2,o}
 \fmf{vanilla,right,tension=0.3,label=$\bullet$,label.dist=0}{v1,v2}
 \fmf{vanilla,left,tension=0.3}{v1,v2}
\end{fmfgraph*}}
\end{fmffile} \hspace{-0.6cm}+\;\;
\left( \hspace{-0.3cm}\begin{fmffile}{fmfq2}\parbox{20mm}{
\begin{fmfgraph*}(12,8)
 \fmftop{i}
 \fmfbottom{o}
 \fmf{vanilla}{i,v,o}
 \fmfdot{i}
\fmfdot{v} \fmfdot{o}
\end{fmfgraph*}}
\end{fmffile}\hspace{-1.1cm}\right)_{IR}\quad
\begin{fmffile}{fmfq30}\parbox{20mm}{
\begin{fmfgraph*}(12,12)
 \fmfleft{i}
 \fmfright{o}
 \fmf{vanilla}{i,v1}
 \fmf{vanilla}{v2,o}
 \fmf{vanilla,left,tension=.3}{v1,v2,v1}
\end{fmfgraph*}} \end{fmffile}\hspace{-0.6cm}
\right)= \\
\\
R_{UV}\left(
\begin{fmffile}{fmfq29}\parbox{20mm}{
\begin{fmfgraph*}(12,12)
 \fmfleft{i}
 \fmfright{o}
 \fmf{vanilla,tension=15}{i,v1,v2,o}
 \fmf{vanilla,right,tension=0.3,label=$\bullet$,label.dist=0}{v1,v2}
 \fmf{vanilla,left,tension=0.3}{v1,v2}
\end{fmfgraph*}}
\end{fmffile} \hspace{-0.6cm}
\right) + \left(
\hspace{-0.3cm}\begin{fmffile}{fmfq2}\parbox{20mm}{
\begin{fmfgraph*}(12,8)
 \fmftop{i}
 \fmfbottom{o}
 \fmf{vanilla}{i,v,o}
 \fmfdot{i}
\fmfdot{v} \fmfdot{o}
\end{fmfgraph*}}
\end{fmffile}\hspace{-1.1cm}\right)_{IR}
\;\; R_{UV}\left(
\begin{fmffile}{fmfq30}\parbox{20mm}{
\begin{fmfgraph*}(12,12)
 \fmfleft{i}
 \fmfright{o}
 \fmf{vanilla}{i,v1}
 \fmf{vanilla}{v2,o}
 \fmf{vanilla,left,tension=.3}{v1,v2,v1}
\end{fmfgraph*}} \end{fmffile} \hspace{-0.6cm}
\right)
\end{array}
\end{equation}
We never subtract IRD or UVD from counter terms, so there is no
``nesting" of the subtraction procedure.

The UVD are easily located.  Only the subgraph
\begin{fmffile}{fmfp2}\parbox{20mm}{
\begin{fmfgraph*}(12,8)
 \fmfleft{i}
 \fmfright{o}
 \fmfdot{o}
 \fmfdot{i}
 \fmf{vanilla,left=0.4}{i,o,i}
\end{fmfgraph*}}
\end{fmffile} \hspace{-0.8cm} is UVD.  Hence
\begin{equation}
\begin{array}{l}
R_{UV}\;\;
\begin{fmffile}{fmfq29}\parbox{20mm}{
\begin{fmfgraph*}(12,12)
 \fmfleft{i}
 \fmfright{o}
 \fmf{vanilla,tension=15}{i,v1,v2,o}
 \fmf{vanilla,right,tension=0.3,label=$\bullet$,label.dist=0}{v1,v2}
 \fmf{vanilla,left,tension=0.3}{v1,v2}
\end{fmfgraph*}}
\end{fmffile} \hspace{-0.7cm}=
\begin{fmffile}{fmfq29}\parbox{20mm}{
\begin{fmfgraph*}(12,12)
 \fmfleft{i}
 \fmfright{o}
 \fmf{vanilla,tension=15}{i,v1,v2,o}
 \fmf{vanilla,right,tension=0.3,label=$\bullet$,label.dist=0}{v1,v2}
 \fmf{vanilla,left,tension=0.3}{v1,v2}
\end{fmfgraph*}}
\end{fmffile} \hspace{-0.7cm}+
\left(\hspace{-0.3cm}
\begin{fmffile}{fmfq47}\parbox{20mm}{
\begin{fmfgraph*}(10,6)
\fmftop{t} \fmfbottom{b} \fmfdot{t} \fmfdot{b}
 \fmf{vanilla,left=0.4,tension=1}{t,b}
\fmf{vanilla,right=0.4,tension=1}{t,b}
\end{fmfgraph*}}
\end{fmffile} \hspace{-1.3cm}\right)_{UV}
\begin{fmffile}{fmfp6}
\parbox{20mm}{
\begin{fmfgraph*}(12,8)
  \fmfleft{i} \fmfright{o} \fmf{vanilla}{i,v,o}
  \fmf{vanilla,left,label=$\bullet$,label.dist=0}{v,v}
\end{fmfgraph*}}
\end{fmffile}\hspace{-0.7cm}+
\left(
\begin{fmffile}{fmfq29}\parbox{20mm}{
\begin{fmfgraph*}(12,12)
 \fmfleft{i}
 \fmfright{o}
 \fmf{vanilla,tension=15}{i,v1,v2,o}
 \fmf{vanilla,right,tension=0.3,label=$\bullet$,label.dist=0}{v1,v2}
 \fmf{vanilla,left,tension=0.3}{v1,v2}
\end{fmfgraph*}}
\end{fmffile} \hspace{-0.7cm}
\right)_{UV} \hspace{-0.6cm}=\;\mbox{UV-finite}\\
\\
R_{UV}\;\;
\begin{fmffile}{fmfq30}\parbox{20mm}{
\begin{fmfgraph*}(12,12)
 \fmfleft{i}
 \fmfright{o}
 \fmf{vanilla}{i,v1}
 \fmf{vanilla}{v2,o}
 \fmf{vanilla,left,tension=.3}{v1,v2,v1}
\end{fmfgraph*}} \end{fmffile}  \hspace{-0.7cm}=
\begin{fmffile}{fmfq30}\parbox{20mm}{
\begin{fmfgraph*}(12,12)
 \fmfleft{i}
 \fmfright{o}
 \fmf{vanilla}{i,v1}
 \fmf{vanilla}{v2,o}
 \fmf{vanilla,left,tension=.3}{v1,v2,v1}
\end{fmfgraph*}} \end{fmffile}  \hspace{-0.7cm}+
\left(\hspace{-0.3cm}
\begin{fmffile}{fmfq47}\parbox{20mm}{
\begin{fmfgraph*}(10,6)
\fmftop{t} \fmfbottom{b} \fmfdot{t} \fmfdot{b}
 \fmf{vanilla,left=0.4,tension=1}{t,b}
\fmf{vanilla,right=0.4,tension=1}{t,b}
\end{fmfgraph*}}
\end{fmffile} \hspace{-1.3cm}\right)_{UV}
\begin{fmffile}{fmfp7}
\parbox{20mm}{
\begin{fmfgraph*}(12,8)
  \fmfleft{i} \fmfright{o}\fmfdot{v} \fmf{vanilla}{i,v,o}
  \end{fmfgraph*}}
\end{fmffile}\hspace{-0.7cm}
=\;\mbox{UV-finite}
\end{array}
\end{equation}
The tadpole graph in the first line vanishes according to the rules of
dimensional regularization.  (We set it
to zero even though it contains an infrared and an ultraviolet
divergence.  The ultraviolet divergence in
this diagram is accounted for by the whole $R^\ast$ procedure).  Note
that again we always begin with
the original graph on the right-hand side, and then subtract divergences
by adding counter terms.  (Again by
convention we use plus signs for these subtractions).

Combining the IR and UV subtraction procedure, we obtain for the
graph in (\ref{zabil}) \begin{equation}
\begin{array}{l}
 R^*\;\; \left(
\begin{fmffile}{fmfq29}\parbox{20mm}{
\begin{fmfgraph*}(12,12)
 \fmfleft{i}
 \fmfright{o}
 \fmf{vanilla,tension=15}{i,v1,v2,o}
 \fmf{vanilla,right,tension=0.3,label=$\bullet$,label.dist=0}{v1,v2}
 \fmf{vanilla,left,tension=0.3}{v1,v2}
\end{fmfgraph*}}
\end{fmffile} \hspace{-0.7cm}
\right) =
\begin{fmffile}{fmfq29}\parbox{20mm}{
\begin{fmfgraph*}(12,12)
 \fmfleft{i}
 \fmfright{o}
 \fmf{vanilla,tension=15}{i,v1,v2,o}
 \fmf{vanilla,right,tension=0.3,label=$\bullet$,label.dist=0}{v1,v2}
 \fmf{vanilla,left,tension=0.3}{v1,v2}
\end{fmfgraph*}}
\end{fmffile} \hspace{-0.7cm}+
\left(\hspace{-0.3cm}
\begin{fmffile}{fmfq47}\parbox{20mm}{
\begin{fmfgraph*}(10,6)
\fmftop{t} \fmfbottom{b} \fmfdot{t} \fmfdot{b}
 \fmf{vanilla,left=0.4,tension=1}{t,b}
\fmf{vanilla,right=0.4,tension=1}{t,b}
\end{fmfgraph*}}
\end{fmffile} \hspace{-1.3cm}\right)_{UV}
\begin{fmffile}{fmfp6}
\parbox{20mm}{
\begin{fmfgraph*}(12,8)
  \fmfleft{i} \fmfright{o} \fmf{vanilla}{i,v,o}
  \fmf{vanilla,left,label=$\bullet$,label.dist=0}{v,v}
\end{fmfgraph*}}
\end{fmffile}\hspace{-0.7cm}+
\left(
\begin{fmffile}{fmfq29}\parbox{20mm}{
\begin{fmfgraph*}(12,12)
 \fmfleft{i}
 \fmfright{o}
 \fmf{vanilla,tension=15}{i,v1,v2,o}
 \fmf{vanilla,right,tension=0.3,label=$\bullet$,label.dist=0}{v1,v2}
 \fmf{vanilla,left,tension=0.3}{v1,v2}
\end{fmfgraph*}}
\end{fmffile} \hspace{-0.7cm}
\right)_{UV} \\
\\
+\left( \hspace{-0.3cm}\begin{fmffile}{fmfq2}\parbox{20mm}{
\begin{fmfgraph*}(12,8)
 \fmftop{i}
 \fmfbottom{o}
 \fmf{vanilla}{i,v,o}
 \fmfdot{i}
\fmfdot{v} \fmfdot{o}
\end{fmfgraph*}}
\end{fmffile}\hspace{-1.1cm}\right)_{IR}\quad
\begin{fmffile}{fmfq30}\parbox{20mm}{
\begin{fmfgraph*}(12,12)
 \fmfleft{i}
 \fmfright{o}
 \fmf{vanilla}{i,v1}
 \fmf{vanilla}{v2,o}
 \fmf{vanilla,left,tension=.3}{v1,v2,v1}
\end{fmfgraph*}} \end{fmffile}\hspace{-0.5cm}+
\;\; \left( \hspace{-0.3cm}\begin{fmffile}{fmfq2}\parbox{20mm}{
\begin{fmfgraph*}(12,8)
 \fmftop{i}
 \fmfbottom{o}
 \fmf{vanilla}{i,v,o}
 \fmfdot{i}
\fmfdot{v} \fmfdot{o}
\end{fmfgraph*}}
\end{fmffile}\hspace{-1.1cm}\right)_{IR}
\left(\hspace{-0.3cm}
\begin{fmffile}{fmfq47}\parbox{20mm}{
\begin{fmfgraph*}(10,6)
\fmftop{t} \fmfbottom{b} \fmfdot{t} \fmfdot{b}
 \fmf{vanilla,left=0.4,tension=1}{t,b}
\fmf{vanilla,right=0.4,tension=1}{t,b}
\end{fmfgraph*}}
\end{fmffile} \hspace{-1.3cm}\right)_{UV}
\begin{fmffile}{fmfp7}
\parbox{20mm}{
\begin{fmfgraph*}(12,8)
  \fmfleft{i} \fmfright{o}\fmfdot{v} \fmf{vanilla}{i,v,o}
  \end{fmfgraph*}}
\end{fmffile}\hspace{-0.7cm}
=\;\mbox{finite}
\end{array}
\end{equation} We can now determine the
overall UV counter term which makes the graph finite after all
subdivergences have been removed.  This is the UV counter term
one needs for the $\beta$ function.  It is given by $(
\begin{fmffile}{fmfq70}\parbox{20mm}{
\begin{fmfgraph*}(10,10)
 \fmfleft{i}
 \fmfright{o}
 \fmf{vanilla,tension=20}{i,v1,v2,o}
 \fmf{vanilla,right,tension=0.3,label=$\bullet$,label.dist=0}{v1,v2}
 \fmf{vanilla,left,tension=0.3}{v1,v2}
\end{fmfgraph*}}
\end{fmffile} \hspace{-1.0cm}
)_{UV}$ and can be
computed by evaluating the r.h.s. of the following equation
\begin{equation}
\begin{array}{l}
\left(
\begin{fmffile}{fmfq29}\parbox{20mm}{
\begin{fmfgraph*}(12,12)
 \fmfleft{i}
 \fmfright{o}
 \fmf{vanilla,tension=15}{i,v1,v2,o}
 \fmf{vanilla,right,tension=0.3,label=$\bullet$,label.dist=0}{v1,v2}
 \fmf{vanilla,left,tension=0.3}{v1,v2}
\end{fmfgraph*}}
\end{fmffile} \hspace{-0.7cm}
\right)_{UV}= - \left[
\begin{fmffile}{fmfq29}\parbox{20mm}{
\begin{fmfgraph*}(12,12)
 \fmfleft{i}
 \fmfright{o}
 \fmf{vanilla,tension=15}{i,v1,v2,o}
 \fmf{vanilla,right,tension=0.3,label=$\bullet$,label.dist=0}{v1,v2}
 \fmf{vanilla,left,tension=0.3}{v1,v2}
\end{fmfgraph*}}
\end{fmffile} \hspace{-0.7cm}+
\left(\hspace{-0.3cm}
\begin{fmffile}{fmfq47}\parbox{20mm}{
\begin{fmfgraph*}(10,6)
\fmftop{t} \fmfbottom{b} \fmfdot{t} \fmfdot{b}
 \fmf{vanilla,left=0.4,tension=1}{t,b}
\fmf{vanilla,right=0.4,tension=1}{t,b}
\end{fmfgraph*}}
\end{fmffile} \hspace{-1.3cm}\right)_{UV}
\begin{fmffile}{fmfp6}
\parbox{20mm}{
\begin{fmfgraph*}(12,8)
  \fmfleft{i} \fmfright{o} \fmf{vanilla}{i,v,o}
  \fmf{vanilla,left,label=$\bullet$,label.dist=0}{v,v}
\end{fmfgraph*}}
\end{fmffile}\hspace{-0.7cm}+\right.
 \\
\\
\left. \left( \hspace{-0.3cm}\begin{fmffile}{fmfq2}\parbox{20mm}{
\begin{fmfgraph*}(12,8)
 \fmftop{i}
 \fmfbottom{o}
 \fmf{vanilla}{i,v,o}
 \fmfdot{i}
\fmfdot{v} \fmfdot{o}
\end{fmfgraph*}}
\end{fmffile}\hspace{-1.1cm}\right)_{IR}\quad
\begin{fmffile}{fmfq30}\parbox{20mm}{
\begin{fmfgraph*}(12,12)
 \fmfleft{i}
 \fmfright{o}
 \fmf{vanilla}{i,v1}
 \fmf{vanilla}{v2,o}
 \fmf{vanilla,left,tension=.3}{v1,v2,v1}
\end{fmfgraph*}} \end{fmffile}\hspace{-0.5cm}+
\;\; \left( \hspace{-0.3cm}\begin{fmffile}{fmfq2}\parbox{20mm}{
\begin{fmfgraph*}(12,8)
 \fmftop{i}
 \fmfbottom{o}
 \fmf{vanilla}{i,v,o}
 \fmfdot{i}
\fmfdot{v} \fmfdot{o}
\end{fmfgraph*}}
\end{fmffile}\hspace{-1.1cm}\right)_{IR}
\left(\hspace{-0.3cm}
\begin{fmffile}{fmfq47}\parbox{20mm}{
\begin{fmfgraph*}(10,6)
\fmftop{t} \fmfbottom{b} \fmfdot{t} \fmfdot{b}
 \fmf{vanilla,left=0.4,tension=1}{t,b}
\fmf{vanilla,right=0.4,tension=1}{t,b}
\end{fmfgraph*}}
\end{fmffile} \hspace{-1.3cm}\right)_{UV}
\right]+\;\mbox{finite parts}
\end{array}
\end{equation} Taking the pole parts (PP), we can
also write \begin{equation}
\begin{array}{l}
\left[ \left(
\begin{fmffile}{fmfq29}\parbox{20mm}{
\begin{fmfgraph*}(12,12)
 \fmfleft{i}
 \fmfright{o}
 \fmf{vanilla,tension=15}{i,v1,v2,o}
 \fmf{vanilla,right,tension=0.3,label=$\bullet$,label.dist=0}{v1,v2}
 \fmf{vanilla,left,tension=0.3}{v1,v2}
\end{fmfgraph*}}
\end{fmffile} \hspace{-0.7cm}
\right)_{UV}\right]_{PP}= - \left[
\begin{fmffile}{fmfq29}\parbox{20mm}{
\begin{fmfgraph*}(12,12)
 \fmfleft{i}
 \fmfright{o}
 \fmf{vanilla,tension=15}{i,v1,v2,o}
 \fmf{vanilla,right,tension=0.3,label=$\bullet$,label.dist=0}{v1,v2}
 \fmf{vanilla,left,tension=0.3}{v1,v2}
\end{fmfgraph*}}
\end{fmffile} \hspace{-0.7cm}+
\left(\hspace{-0.3cm}
\begin{fmffile}{fmfq47}\parbox{20mm}{
\begin{fmfgraph*}(10,6)
\fmftop{t} \fmfbottom{b} \fmfdot{t} \fmfdot{b}
 \fmf{vanilla,left=0.4,tension=1}{t,b}
\fmf{vanilla,right=0.4,tension=1}{t,b}
\end{fmfgraph*}}
\end{fmffile} \hspace{-1.3cm}\right)_{UV}
\begin{fmffile}{fmfp6}
\parbox{20mm}{
\begin{fmfgraph*}(12,8)
  \fmfleft{i} \fmfright{o} \fmf{vanilla}{i,v,o}
  \fmf{vanilla,left,label=$\bullet$,label.dist=0}{v,v}
\end{fmfgraph*}}
\end{fmffile}\hspace{-0.7cm}+\right.
 \\
\\
\left. \left( \hspace{-0.3cm}\begin{fmffile}{fmfq2}\parbox{20mm}{
\begin{fmfgraph*}(12,8)
 \fmftop{i}
 \fmfbottom{o}
 \fmf{vanilla}{i,v,o}
 \fmfdot{i}
\fmfdot{v} \fmfdot{o}
\end{fmfgraph*}}
\end{fmffile}\hspace{-1.1cm}\right)_{IR}\quad
\begin{fmffile}{fmfq30}\parbox{20mm}{
\begin{fmfgraph*}(12,12)
 \fmfleft{i}
 \fmfright{o}
 \fmf{vanilla}{i,v1}
 \fmf{vanilla}{v2,o}
 \fmf{vanilla,left,tension=.3}{v1,v2,v1}
\end{fmfgraph*}} \end{fmffile}\hspace{-0.5cm}+
\;\; \left( \hspace{-0.3cm}\begin{fmffile}{fmfq2}\parbox{20mm}{
\begin{fmfgraph*}(12,8)
 \fmftop{i}
 \fmfbottom{o}
 \fmf{vanilla}{i,v,o}
 \fmfdot{i}
\fmfdot{v} \fmfdot{o}
\end{fmfgraph*}}
\end{fmffile}\hspace{-1.1cm}\right)_{IR}
\left(\hspace{-0.3cm}
\begin{fmffile}{fmfq47}\parbox{20mm}{
\begin{fmfgraph*}(10,6)
\fmftop{t} \fmfbottom{b} \fmfdot{t} \fmfdot{b}
 \fmf{vanilla,left=0.4,tension=1}{t,b}
\fmf{vanilla,right=0.4,tension=1}{t,b}
\end{fmfgraph*}}
\end{fmffile} \hspace{-1.3cm}\right)_{UV}
\right]_{PP}
\end{array}
\end{equation} Each graph on the r.h.s. is
computed with dimensional regularization, including the original
graph (which is of course the most difficult to compute).  The
subtraction terms denoted by $()_{UV}$ and $()_{IR}$ are
polynomials in ${1 \over \e}$, so one must compute also some
graphs on the right-hand side to order $\e, \e^2$ etc.  More
precisely when there is a higher order pole ${1 \over \e^k}$ due
to $()_{UV}$ and $()_{IR}$, one must compute the corresponding
graph to order $\e^{k-1}$.

To illustrate the analogies and differences of $R_{UV}$ and
$R_{IR}$ consider the following example
\begin{equation}
\includegraphics
[draft=false, height=2in,width=5in,keepaspectratio] {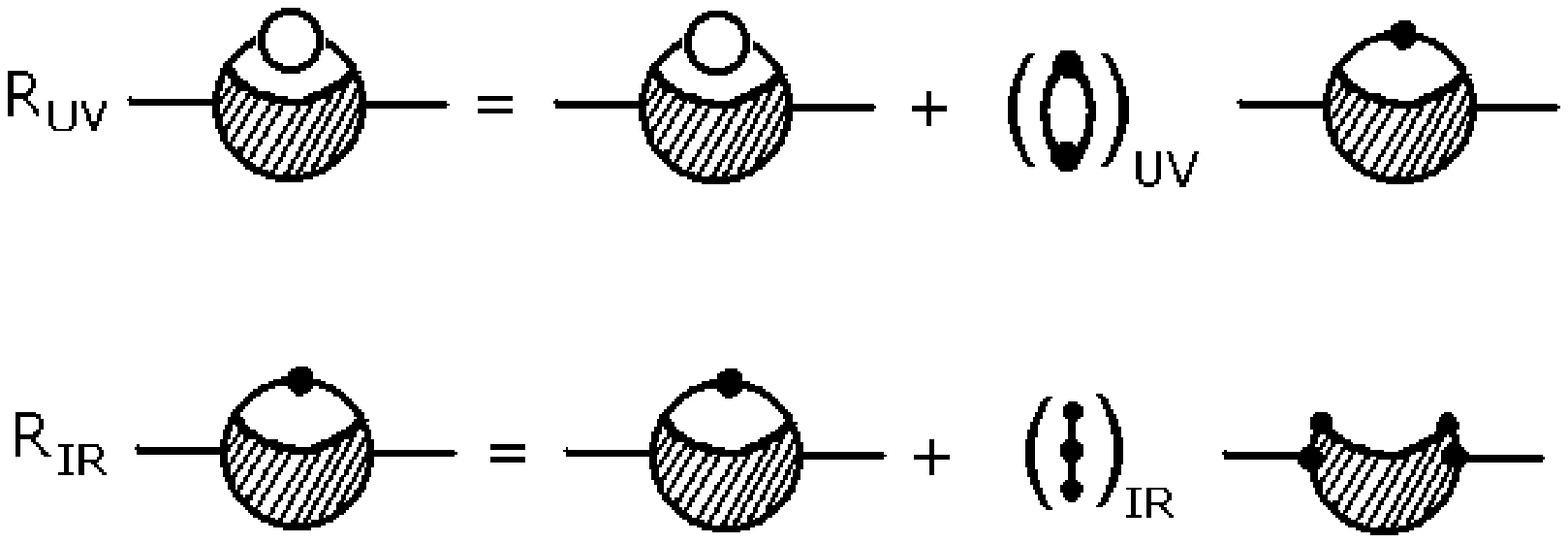}
\end{equation}
The hatched part of the graph denotes any
nonsingular subgraph.  As this example shows, to remove UVD one
shrinks the divergent proper subgraphs to a point, but one deletes
the IR divergent subgraphs.  We compute $\left(\hspace{-0.3cm}
\begin{fmffile}{fmfq47}\parbox{20mm}{
\begin{fmfgraph*}(10,6)
\fmftop{t} \fmfbottom{b} \fmfdot{t} \fmfdot{b}
 \fmf{vanilla,left=0.4,tension=1}{t,b}
\fmf{vanilla,right=0.4,tension=1}{t,b}
\end{fmfgraph*}}
\end{fmffile} \hspace{-1.3cm}\right)_{UV}$ as the ${1 \over \e}$ pole
part of this one-loop graph, and $\left(
\hspace{-0.3cm}\begin{fmffile}{fmfq2}\parbox{20mm}{
\begin{fmfgraph*}(12,8)
 \fmftop{i}
 \fmfbottom{o}
 \fmf{vanilla}{i,v,o}
 \fmfdot{i}
\fmfdot{v} \fmfdot{o}
\end{fmfgraph*}}
\end{fmffile}\hspace{-1.1cm}\right)_{IR}$ is computed from 
\begin{equation}
\begin{array}{l}
R_{IR}\;\;
\begin{fmffile}{fmfq88888}\parbox{20mm}{
\begin{fmfgraph*}(12,12)
 \fmfleft{i}
 \fmfright{o}
 \fmf{vanilla}{i,v1}
 \fmf{vanilla}{v2,o}
\fmf{vanilla,left,tension=0.3,label=$\bullet$,label.dist=0}{v1,v2}
\fmf{vanilla,right,tension=0.3}{v1,v2}
\end{fmfgraph*}}
\end{fmffile}\hspace{-0.6cm}=\
\begin{fmffile}{fmfq88888}\parbox{20mm}{
\begin{fmfgraph*}(12,12)
 \fmfleft{i}
 \fmfright{o}
 \fmf{vanilla}{i,v1}
 \fmf{vanilla}{v2,o}
\fmf{vanilla,left,tension=0.3,label=$\bullet$,label.dist=0}{v1,v2}
\fmf{vanilla,right,tension=0.3}{v1,v2}
\end{fmfgraph*}}
\end{fmffile}
\hspace{-0.6cm}+\;
\left( \hspace{-0.3cm}\begin{fmffile}{fmfq2}\parbox{20mm}{
\begin{fmfgraph*}(12,8)
 \fmftop{i}
 \fmfbottom{o}
 \fmf{vanilla}{i,v,o}
 \fmfdot{i}
\fmfdot{v} \fmfdot{o}
\end{fmfgraph*}}
\end{fmffile}\hspace{-1.1cm}\right)_{IR}
\begin{fmffile}{fmfq222}\parbox{20mm}{
\begin{fmfgraph*}(12,8)
 \fmfleft{i}
 \fmfright{o}
 \fmf{vanilla}{i,v1,v2,o}
 \fmfdot{v2}
\fmfdot{v1}
\end{fmfgraph*}}
\end{fmffile} \hspace{-0.6cm} = \; \mbox{finite}\;\; \left(\mbox{no}\;\frac{1}{\epsilon} \right)
\end{array}
\end{equation} In
both cases one gets ${1 \over \e}$ poles (if one uses dimensional
regularization).

Recall that the graphical identity
\begin{equation}
\begin{array}{l}
R_{IR}\;\;
\begin{fmffile}{fmfq88888}\parbox{20mm}{
\begin{fmfgraph*}(12,12)
 \fmfleft{i}
 \fmfright{o}
 \fmf{vanilla}{i,v1}
 \fmf{vanilla}{v2,o}
\fmf{vanilla,left,tension=0.3,label=$\bullet$,label.dist=0}{v1,v2}
\fmf{vanilla,right,tension=0.3}{v1,v2}
\end{fmfgraph*}}
\end{fmffile}
\hspace{-0.6cm}=\;\;
\begin{fmffile}{fmfq88888}\parbox{20mm}{
\begin{fmfgraph*}(12,12)
 \fmfleft{i}
 \fmfright{o}
 \fmf{vanilla}{i,v1}
 \fmf{vanilla}{v2,o}
\fmf{vanilla,left,tension=0.3,label=$\bullet$,label.dist=0}{v1,v2}
\fmf{vanilla,right,tension=0.3}{v1,v2}
\end{fmfgraph*}}
\end{fmffile}\hspace{-0.6cm}+\;
\left( \hspace{-0.3cm}\begin{fmffile}{fmfq2}\parbox{20mm}{
\begin{fmfgraph*}(12,8)
 \fmftop{i}
 \fmfbottom{o}
 \fmf{vanilla}{i,v,o}
 \fmfdot{i}
\fmfdot{v} \fmfdot{o}
\end{fmfgraph*}}
\end{fmffile}\hspace{-1.1cm}\right)_{IR}
\begin{fmffile}{fmfq222}\parbox{20mm}{
\begin{fmfgraph*}(12,8)
 \fmfleft{i}
 \fmfright{o}
 \fmf{vanilla}{i,v1,v2,o}
 \fmfdot{v2}
\fmfdot{v1}
\end{fmfgraph*}}
\end{fmffile}
\end{array}
\end{equation}
corresponds to the following analytical expression \eqa R_{IR}
\int d^4 k {1 \over k^4} {1 \over (k-Q)^2} = \int d^4 k \left( {1
\over k^4} + {c \over \e} \delta^4 (k) \right) {1 \over (k-Q)^2}
\eqae The IRD occurs at $k=0$ and is taken care of by the ${1
\over \e} \delta^4 (k)$ insertion.  One may find products of such
IR factors $\delta^4 (k) / \e$ with different loop momenta, but
never with the same loop momentum.  As an example consider the
following Feynman graph in $\varphi^4 + \varphi^3$ theory
\eqa
\begin{fmffile}{fmfq103}\parbox{20mm}{
\begin{fmfgraph*}(40,30)
\fmfleft{i} \fmfright{o} \fmftop{t} \fmfbottom{b}
\fmf{fermion,tension=5,label=$Q$}{v1,i}
\fmf{fermion,tension=5,label=$Q$}{o,v2}
\fmf{fermion,label=$p$,l.side=right}{t,v1}
\fmf{fermion,label=$k$}{v2,t}
\fmf{fermion,label=$Q-p$,left=0.45,tension=0}{b,v1}
\fmf{fermion,label=$p-q$,l.side=right,label.dist=1.1mm,label.angle=-30,left=0.45,tension=0}{v1,b}
\fmf{fermion,label=$Q-k$}{v2,b} \fmf{vanilla}{t,v}
 \fmfdot{v}
\fmf{fermion,label=$k-p$,l.side=left}{v,v4} \fmf{vanilla}{v4,b}
 \fmf{phantom}{b,v1}
\end{fmfgraph*}}
\end{fmffile} \qquad\qquad\qquad F = \int
{d^4 kd^4 q d^4 p \over (k-p)^4 p^2 k^2} {1 \over (Q-k)^2} {1
\over (Q -q)^2 (p-q)^2} \eqae In this example there is an IRD at
$k=p$, and an overall IRD at $k = p=0$.  To subtract these IRD we
replace some propagators by delta function in $D$ dimensions
\begin{equation}
\begin{array}{l}
R_{IR} (F) = F + \int \left( {1 \over (k-p)^4} \rightarrow \mu^\e
C_1 \delta^D (k-p) \right) {1 \over p^2} {1 \over k^2} {1 \over
(Q-k)^2} {1 \over (Q-q)^2} {1 \over (p-q)^2} \\
+ \int \left( {1 \over (p-k)^4} {1 \over p^2} {1 \over k^2}
\rightarrow (\mu^{2 \e})^2 C_2 \delta^D (p) \delta^D (k) \right)
\left ( {1 \over (Q-k)^2 (Q-q)^2} {1 \over (p-q)^2} \right)
\end{array}
\end{equation}
Because there are now two $D$-dimensional Dirac functions one
needs the factor $\mu^{2 \e}$ with $\e = D - 4$, as usual in
dimensional regularization, to make the dimensions come out
correctly.

We can write this pictorially as follows
\begin{equation}
\begin{array}{l}
 R_{IR}\;\; \left(
\begin{fmffile}{fmfr2}\parbox{20mm}{
\begin{fmfgraph*}(15,12)
\fmfleft{i} \fmfright{o} \fmftop{t} \fmfbottom{b}
\fmf{vanilla,tension=5}{v1,i} \fmf{vanilla,tension=5}{o,v2}
\fmf{vanilla}{t,v1} \fmf{vanilla}{v2,t}
\fmf{vanilla,left=0.4,tension=0}{b,v1,b} \fmf{vanilla}{v2,b}
\fmf{vanilla}{t,v,b}
 \fmfdot{v}
 \fmf{phantom}{b,v1}
\end{fmfgraph*}}
\end{fmffile}\hspace{-0.4cm}
\right) =\;
\begin{fmffile}{fmfr2}\parbox{20mm}{
\begin{fmfgraph*}(15,12)
\fmfleft{i} \fmfright{o} \fmftop{t} \fmfbottom{b}
\fmf{vanilla,tension=5}{v1,i} \fmf{vanilla,tension=5}{o,v2}
\fmf{vanilla}{t,v1} \fmf{vanilla}{v2,t}
\fmf{vanilla,left=0.4,tension=0}{b,v1,b} \fmf{vanilla}{v2,b}
\fmf{vanilla}{t,v,b}
 \fmfdot{v}
 \fmf{phantom}{b,v1}
\end{fmfgraph*}}
\end{fmffile}\hspace{-0.3cm}+\;
\left( \hspace{-0.3cm}\begin{fmffile}{fmfq2}\parbox{20mm}{
\begin{fmfgraph*}(12,8)
 \fmftop{i}
 \fmfbottom{o}
 \fmf{vanilla}{i,v,o}
 \fmfdot{i}
\fmfdot{v} \fmfdot{o}
\end{fmfgraph*}}
\end{fmffile}\hspace{-1.1cm}\right)_{IR}
\begin{fmffile}{fmfr3}\parbox{20mm}{
\begin{fmfgraph*}(15,12)
\fmfleft{i} \fmfright{o} \fmftop{t} \fmfbottom{b}
\fmf{vanilla,tension=5}{v1,i} \fmf{vanilla,tension=5}{o,v2}
\fmf{vanilla}{t,v1} \fmf{vanilla}{v2,t}
\fmf{vanilla,left=0.4,tension=0}{b,v1,b} \fmf{vanilla}{v2,b}
 \fmf{phantom}{b,v1}
\end{fmfgraph*}}
\end{fmffile}\\
\\
+\; \left(
\begin{fmffile}{fmfr4}\parbox{20mm}{
	\begin{fmfgraph*}(15,12)
\fmfleft{i} \fmfright{o} \fmftop{t} \fmfbottom{b}
\fmf{vanilla}{i,t,o} \fmf{vanilla}{b,v,t} \fmfdot{t} \fmfdot{i}
\fmfdot{v} \fmfdot{b} \fmfdot{o}
\end{fmfgraph*}}
\end{fmffile}\hspace{-0.4cm}
\right)_{IR}\hspace{-0.2cm}
\begin{fmffile}{fmfr5}\parbox{20mm}{
\begin{fmfgraph*}(12,12)
 \fmfleft{i}
 \fmfright{o}
 \fmf{vanilla}{i,v1}
 \fmf{vanilla}{v2,v3,o}
 \fmfdot{v3}
 \fmf{vanilla,left,tension=.3}{v1,v2,v1}
\end{fmfgraph*}} \end{fmffile}\hspace{-0.4cm} = \;\; \mbox{IR-finite}
\end{array}
\end{equation}
We recall that we must first compute $C_1$ and $C_2$, and then by
substitution we should find that the r.h.s. is IR-finite.

The computation of $C_1$ was discussed before, it follows from
$\hspace{0.1cm}
\begin{fmffile}{fmfq7777}\parbox{20mm}{
\begin{fmfgraph*}(12,12)
 \fmfleft{i}
 \fmfright{o}
 \fmf{vanilla}{i,v1}
 \fmf{vanilla}{v2,o}
\fmf{vanilla,right,tension=0.3,label=$\bullet$,label.dist=0}{v1,v2}
\fmf{vanilla,left,tension=0.3}{v1,v2}
\end{fmfgraph*}}
\end{fmffile}\hspace{-0.5cm}.
$ The computation of
$C_2$ follows from requiring that the following expression be
IR-finite \begin{equation}
\begin{array}{l}
 R_{IR}\;\; \left(
\begin{fmffile}{fmfr7}\parbox{20mm}{
\begin{fmfgraph*}(15,12)
\fmfleft{i} \fmfright{o} \fmftop{t} \fmfbottom{b}
\fmf{vanilla,tension=5}{v1,i} \fmf{vanilla,tension=5}{o,v2}
\fmf{vanilla}{t,v1} \fmf{vanilla}{v2,t}
\fmf{vanilla}{v2,b} \fmf{vanilla}{t,v,b}
 \fmfdot{v}
 \fmf{vanilla}{b,v1}
\end{fmfgraph*}}
\end{fmffile}\hspace{-0.4cm}
\right) =\;
\begin{fmffile}{fmfr7}\parbox{20mm}{
\begin{fmfgraph*}(15,12)
\fmfleft{i} \fmfright{o} \fmftop{t} \fmfbottom{b}
\fmf{vanilla,tension=5}{v1,i} \fmf{vanilla,tension=5}{o,v2}
\fmf{vanilla}{t,v1} \fmf{vanilla}{v2,t}
\fmf{vanilla}{v2,b} \fmf{vanilla}{t,v,b}
 \fmfdot{v}
 \fmf{vanilla}{b,v1}
\end{fmfgraph*}}
\end{fmffile}\hspace{-0.3cm}+\;
\left( \hspace{-0.3cm}\begin{fmffile}{fmfq2}\parbox{20mm}{
\begin{fmfgraph*}(12,8)
 \fmftop{i}
 \fmfbottom{o}
 \fmf{vanilla}{i,v,o}
 \fmfdot{i}
\fmfdot{v} \fmfdot{o}
\end{fmfgraph*}}
\end{fmffile}\hspace{-1.1cm}\right)_{IR}
\begin{fmffile}{fmfr8}\parbox{20mm}{
\begin{fmfgraph*}(15,12)
\fmfleft{i} \fmfright{o} \fmftop{t} \fmfbottom{b}
\fmf{vanilla,tension=5}{v1,i} \fmf{vanilla,tension=5}{o,v2}
\fmf{vanilla}{t,v1} \fmf{vanilla}{v2,t}
\fmf{vanilla}{v2,b}
 \fmf{vanilla}{b,v1}
\end{fmfgraph*}}
\end{fmffile}\\
\\
+2\; \left(
\begin{fmffile}{fmfr4}\parbox{20mm}{
\begin{fmfgraph*}(15,12)
\fmfleft{i} \fmfright{o} \fmftop{t} \fmfbottom{b}
\fmf{vanilla}{i,t,o} \fmf{vanilla}{b,v,t} \fmfdot{t} \fmfdot{i}
\fmfdot{v} \fmfdot{b} \fmfdot{o}
\end{fmfgraph*}}
\end{fmffile}\hspace{-0.4cm}
\right)_{IR}\hspace{-0.2cm}
\begin{fmffile}{fmfr9}\parbox{20mm}{
\begin{fmfgraph*}(12,12)
 \fmfleft{i}
 \fmfright{o}
 \fmf{vanilla}{i,v1,v2,v3,o}
 \fmfdot{v1}
\fmfdot{v2} \fmfdot{v3}
\end{fmfgraph*}} \end{fmffile}\hspace{-0.4cm} = \;\; \mbox{IR-finite}
\end{array}
\end{equation} The factor 2 is due
to the two overall IRD, one upstairs and one downstairs.  Note
that $\left(\;
\begin{fmffile}{fmfr10}\parbox{20mm}{
\begin{fmfgraph*}(8,8)
\fmfleft{i} \fmfright{o} \fmf{vanilla}{i,o} \fmfdot{i} \fmfdot{o}
\end{fmfgraph*}}
\end{fmffile}\hspace{-1cm}
\right)_{IR} = {a \over \e}$ but $\left(\;
\begin{fmffile}{fmfr11}\parbox{20mm}{
\begin{fmfgraph*}(8,8)
\fmfleft{i} \fmfright{o} \fmftop{t} \fmfbottom{b}
\fmf{vanilla}{i,t,o} \fmf{vanilla}{b,v,t} \fmfdot{t} \fmfdot{i}
\fmfdot{v} \fmfdot{b} \fmfdot{o}
\end{fmfgraph*}}
\end{fmffile}\hspace{-1cm}
\right)_{IR} = {b \over \e} + {c \over \e^2}$.   (If the two
lower propagators would have been massive, we would have needed a
factor one).  Note also that $\begin{fmffile}{fmfr9}\parbox{20mm}{
\begin{fmfgraph*}(12,12)
 \fmfleft{i}
 \fmfright{o}
 \fmf{vanilla}{i,v1,v2,v3,o}
 \fmfdot{v1}
\fmfdot{v2} \fmfdot{v3}
\end{fmfgraph*}} \end{fmffile}\hspace{-1cm}\; = 1/Q^4$.  Having determined $C_1$ and $C_2$ we can
determine $R_{IR}\;\;
\begin{fmffile}{fmfr12}\parbox{20mm}{
\begin{fmfgraph*}(8,8)
\fmfleft{i} \fmfright{o} \fmftop{t} \fmfbottom{b}
\fmf{vanilla,tension=5}{v1,i} \fmf{vanilla,tension=5}{o,v2}
\fmf{vanilla}{t,v1} \fmf{vanilla}{v2,t}
\fmf{vanilla,left=0.4,tension=0}{b,v1,b} \fmf{vanilla}{v2,b}
\fmf{vanilla}{t,v,b}
 \fmfdot{v}
 \fmf{phantom}{b,v1}
\end{fmfgraph*}}
\end{fmffile}\hspace{-0.7cm}.$

The IR counter terms are always polynomials in ${1 \over \e}$ and
$\delta^4 (k_j)$.  However, as our next example shows, one
sometimes needs derivatives of $\delta^4 (k)$.  Consider the
following massless graph \setlength{\columnsep}{-200pt}
\begin{multicols}{2}
\includegraphics
[draft=false, height=1.5in,width=2in,keepaspectratio] {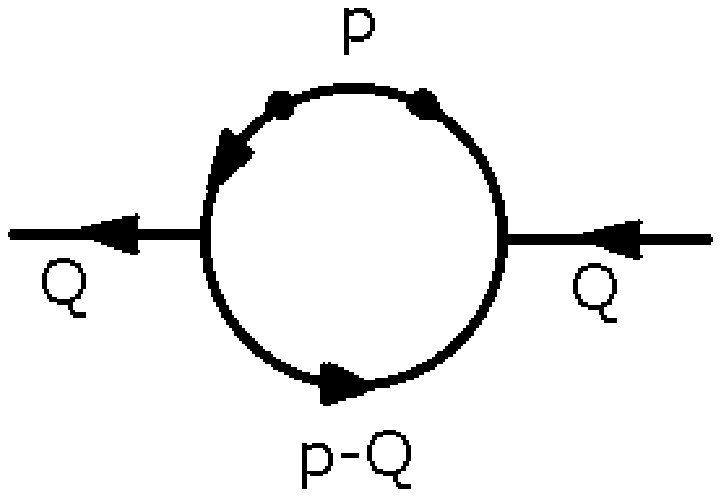}
\\
\\

\begin{equation}
= \int {d^D p \over p^6 (p-Q)^2}
\end{equation}
\end{multicols}

There is now a quadratic IRD proportional to $\int d^4 p /p^6$.
We want to extract it from the original graph by adding a term
involving $\delta^D (p)$ but in order that dimensions match, we
need a Dalembertian acting on $\delta^D (p)$ \eqa\label{dell}
R_{IR} \int {d^D p \over p^6 (p-Q)^2} = \int d^D p  \left[ {1
\over p^6} + \mu^\e \bar{C}_1 \square_p \delta^D (p) \right] {1
\over (p-Q)^2} \eqae Partially integrating the operator
$\square_p = {\del \over \del p^\mu} {\del \over \del p_\mu}$ we
obtain for the last term in the integrand \eqa \int d^D p
\delta^D (p) \square_p {1 \over (p-Q)^2} = \left[ \square_p {1
\over (p-Q)^2} \right] \Bigg|_{p=0} \eqae To fix $\bar{C}_1$ we
consider the simplest graph with this divergence; this is
unfortunately the graph itself but we nevertheless proceed (we
could make the original graph more complicated to lift this
degeneracy).  Hence we fix $\bar{C}_1$ from
\begin{equation}
\includegraphics
[draft=false, height=0.8in,width=5in,keepaspectratio]
{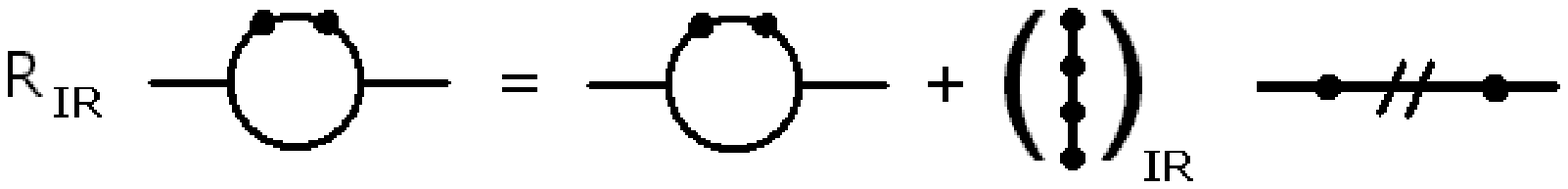}
\end{equation}
The two slashes in the last part of the equation indicate the
action of $\square_p$ on $1/(p-Q)^2$. However, $\square_p
(p-Q)^{-2} \Big|_{p=0} = 2 \e Q^{-4}$, so we find the equation $\;
\begin{fmffile}{fmfr22}\parbox{20mm}{
\begin{fmfgraph*}(12,8)
\fmfcmd{%
   vardef bar (expr p, len, ang) =
    ((-len/2,0)--(len/2,0))
       rotated (ang + angle
      direction length(p)/1 of p)
       shifted point length(p)/1 of p
   enddef;
   style_def tvanilla expr p =
   draw_vanilla p;
  ccutdraw bar (p, 5mm, 70)
  enddef;}
\fmfleft{i} \fmfright{o} \fmf{vanilla}{i,v1}
  \fmf{tvanilla}{v1,v2}
\fmf{vanilla}{o,v4} \fmf{tvanilla}{v4,v3}
  \fmf{vanilla,tension=3}{v2,v3}
  \fmfdot{v1}
  \fmfdot{v4}
\end{fmfgraph*}}
\end{fmffile}\hspace{-0.8cm} \;=\; - 2\epsilon\;\;
\begin{fmffile}{fmfr23}\parbox{20mm}{
\begin{fmfgraph*}(12,8)
\fmfleft{i} \fmfright{o} \fmf{vanilla}{i,v1,v2,v3,o} \fmfdot{v1}
\fmfdot{v2} \fmfdot{v3}
\end{fmfgraph*}}
\end{fmffile}\hspace{-0.8cm}\;
$ and \eqa\label{pit} \int d^D p {1 \over p^6} {1 \over (p-Q)^2} +
\bar{C}_1 2 \e  {1 \over Q^4} = \mbox{IR-finite} \eqae One would
expect that $\bar{C}_1$ is proportional to ${1 \over \e}$, so the
original graph happens to be IR-finite
  (due to the peculiar properties of dimensional regularization).

To evaluate $\bar{C}_1$ we therefore need another graph.  One
could take a massive propagator with $(p-Q)^2$ in which case \eqa
{\square}_p [(p-Q)^2 + m^2]^{-1} = {(8 -2D) (p-Q)^2 -2 D m^2 \over
[(p-Q)^2 + m^2 ]^3} \eqae is no longer proportional to $\e$.  One
can then determine $\bar{C}_1$.

We now discuss a subtlety having to do with the order in which one
applies $R_{UV}$ and $R_{IR}$.  The combined operation is denoted
by $R^\ast$.  A priori one might expect that $R_{IR} R_{UV}$ is
equal to $R_{UV} R_{IR}$, but there exist counter examples at the
five-loop level~\cite{chety}. Consider the following 5-loop graph
\begin{equation}
\includegraphics
[draft=false, height=2in,width=5in,keepaspectratio] {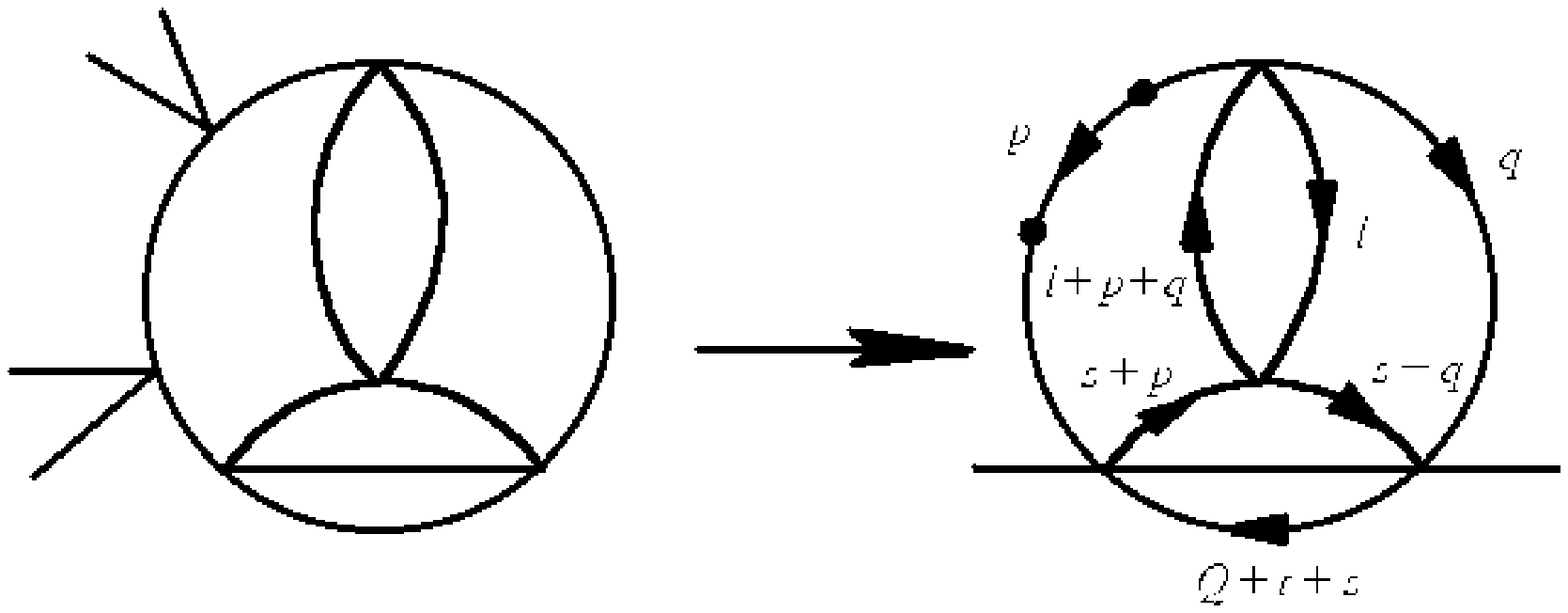}
\end{equation}
Operating with $R_{UV}R_{IR}$ gives the incorrect result
\vspace{5cm}
\begin{equation}
\includegraphics
[draft=false, height=6.5in,width=5in,keepaspectratio]
{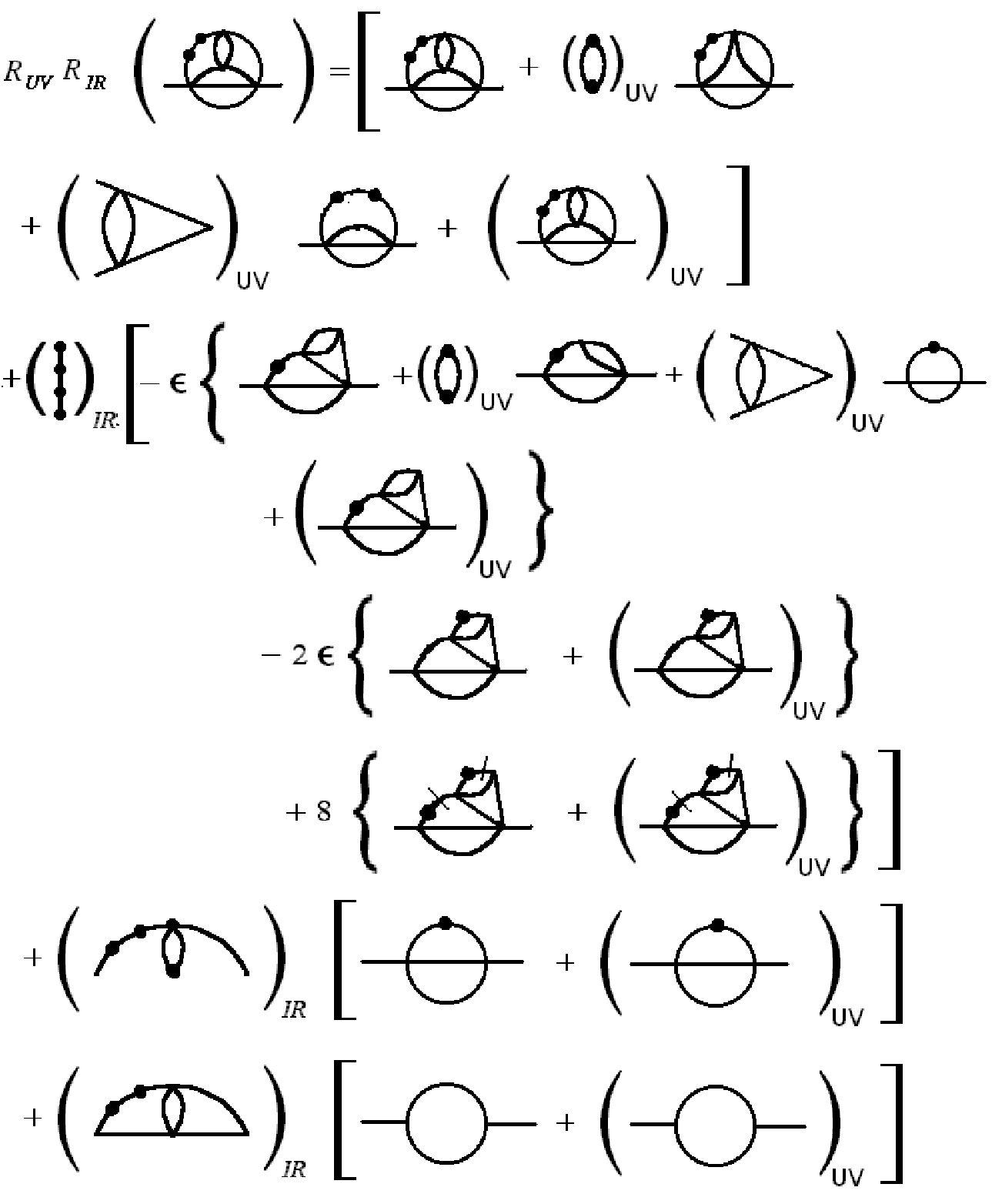}
\end{equation}
The IRD arise when the following loop momenta vanish: $p$, $pql$,
and $pqls$. When $pt$ vanish (or $pl$), one finds an IRD of the
term $\int dpdt p^{-6}t^{-2}$, but this was already considered in
the IRD with $\int dp p^{-6}$ so we do not count it separately.
We repeatedly used that tadpoles vanish, no matter how many loops
they contain. We also used~(\ref{dell})--(\ref{pit}), but because
there are two propagators on which $\square_p$ can act, we get
also cross terms where each propagator carries one derivative.
Operating with $R_{IR}R_{UV}$ gives the correct result
\begin{equation}
\includegraphics
[draft=false, height=6in,width=5.5in,keepaspectratio] {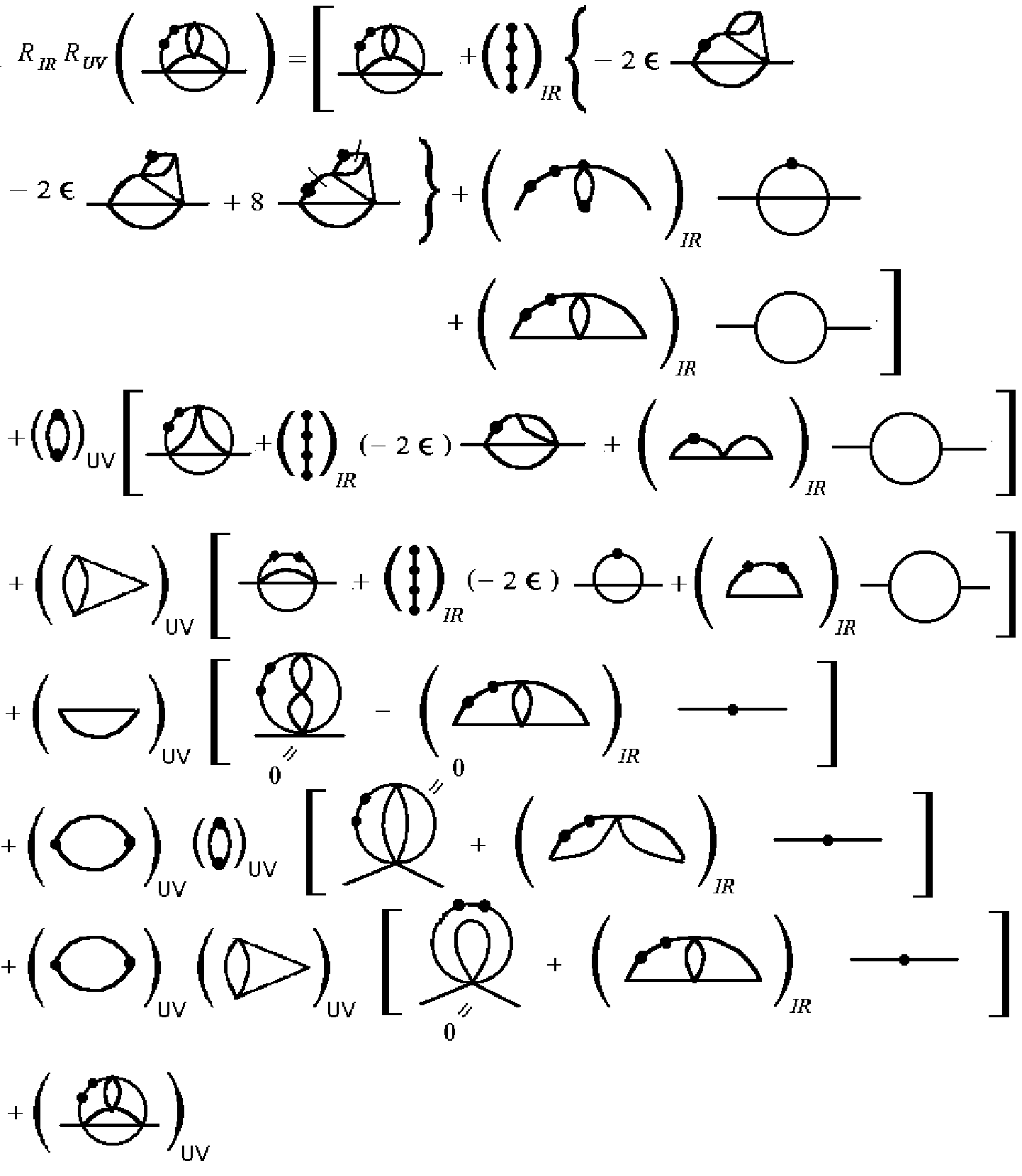}
\end{equation}
 As another example, we consider a 5-loop graph which is needed to
compute the $\beta$ function at 5-loops.  The original diagram
gives a vertex correction for the $\l \varphi^4$ coupling, namely
a graph with 4 external lines, but we nullify all 4 lines.  Since
vacuum graphs vanish in dimensional regularization we add two new
external lines carrying a new external momentum $Q$.
\begin{equation}
\includegraphics
[draft=false, height=2in,width=5.5in,keepaspectratio] {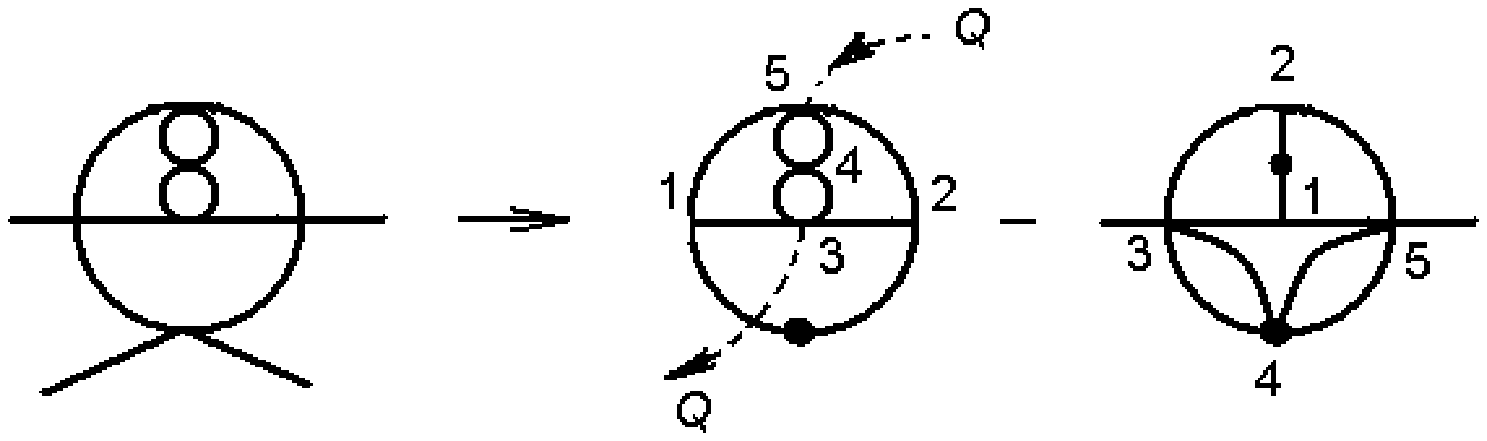}
\end{equation}
The reason we let momenta $Q$ flow in and out the graph at
these particular points is that this allows to compute the
original graph easily. Indeed, to compute the graph itself (which
is always needed), one may first compute the subgraphs
\begin{equation}
\includegraphics
[draft=false, height=0.8in,width=5.5in,keepaspectratio] {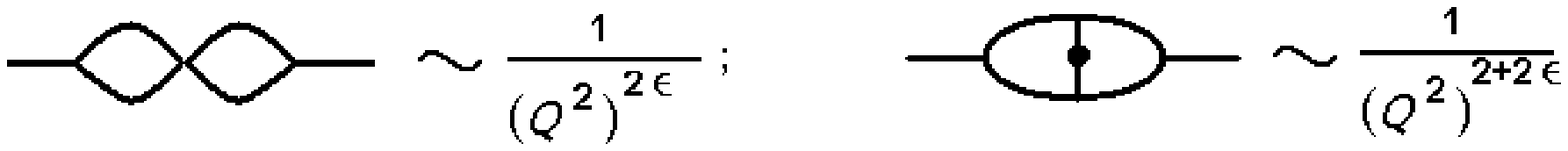}
\end{equation}
Then one evaluates $\int d^D q (Q-q)^2)^{-2 \e} (q^2)^{-2-2
\e}$.  We now evaluate $R^\ast$ on this diagram. Since there are
no ${1 \over p^6}$ terms, there is no term with $\square_p
\delta^D (p)$ and thus no ambiguity whether one should choose
$R_{UV} R_{IR}$ or $R_{IR} R_{UV}$.  We choose the former.  We
first record the result for acting with $R_{IR}$ on the graph,
and then record the results due to acting with $R_{UV}$ on each
of the terms in the result for $R_{IR}$.  We write the results
such that each column in the result for $R_{UV}$ corresponds to
one term in the result for $R_{IR}$.
\includegraphics
[draft=false, height=2in,width=5.5in,keepaspectratio] {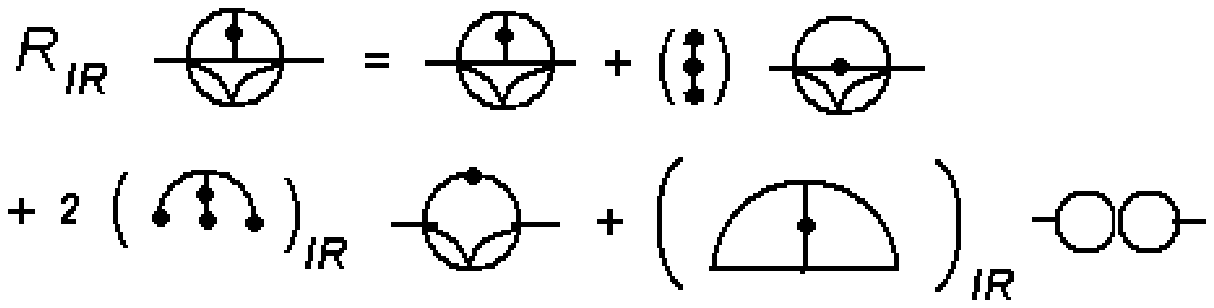}
The subgraph $\left(\;\;
\begin{fmffile}{fmfpp4}\parbox{20mm}{
\begin{fmfgraph*}(15,8)
\fmfstraight \fmfbottom{i,v2,o}
 \fmftop{t}
 \fmf{vanilla}{i,t,o}
 \fmf{vanilla}{t,v1,v2}
 \fmfdot{i}
 \fmfdot{o}
  \fmfdot{i}
  \fmfdot{v1}
   \fmfdot{v2}
\end{fmfgraph*}} \end{fmffile}
\hspace{-0.3cm}\right)_{IR}$ (which corresponds to the IR
divergent integral $\int d^4 k d^4 p (k-p)^{-2} k^{-2} p^{-4}$
for small $k$ and $p$) is written as
$c\delta^{D}(k)\delta^{D}(p)$  whereas the subgraph $\left(\;\;
\begin{fmffile}{fmfpp5}\parbox{20mm}{
\begin{fmfgraph*}(15,8)
\fmfstraight \fmfbottom{i,v2,o}
 \fmftop{t}
 \fmf{vanilla}{i,t,o}
 \fmf{vanilla}{t,v1,v2}
   \fmfdot{v1}
\fmf{vanilla}{i,v2,o}
\end{fmfgraph*}} \end{fmffile}
\hspace{-0.3cm}\right)_{IR}$ contains only one delta function
$\delta^{D}(k+q)$ (since there is only one external momentum, one
can only use one $\delta^D (p)$).  At the end no IRD are left
(\;$\begin{fmffile}{fmfq48}\parbox{20mm}{
\begin{fmfgraph*}(12,12)
 \fmfleft{i}
 \fmfright{o}
 \fmf{vanilla}{i,v1}
 \fmf{vanilla}{v3,o}
 \fmf{vanilla,left,tension=.3}{v1,v2,v1}
\fmf{vanilla,left,tension=.3}{v2,v3,v2}
\end{fmfgraph*}} \end{fmffile}
\hspace{-0.7cm}$ is IR finite).

Now we perform $R_{UV}$ on each of these terms.  We write the result of
each $R_{UV}$ operation as a column.
\eqa
\includegraphics
[draft=false, height=2.5in,width=5in,keepaspectratio] {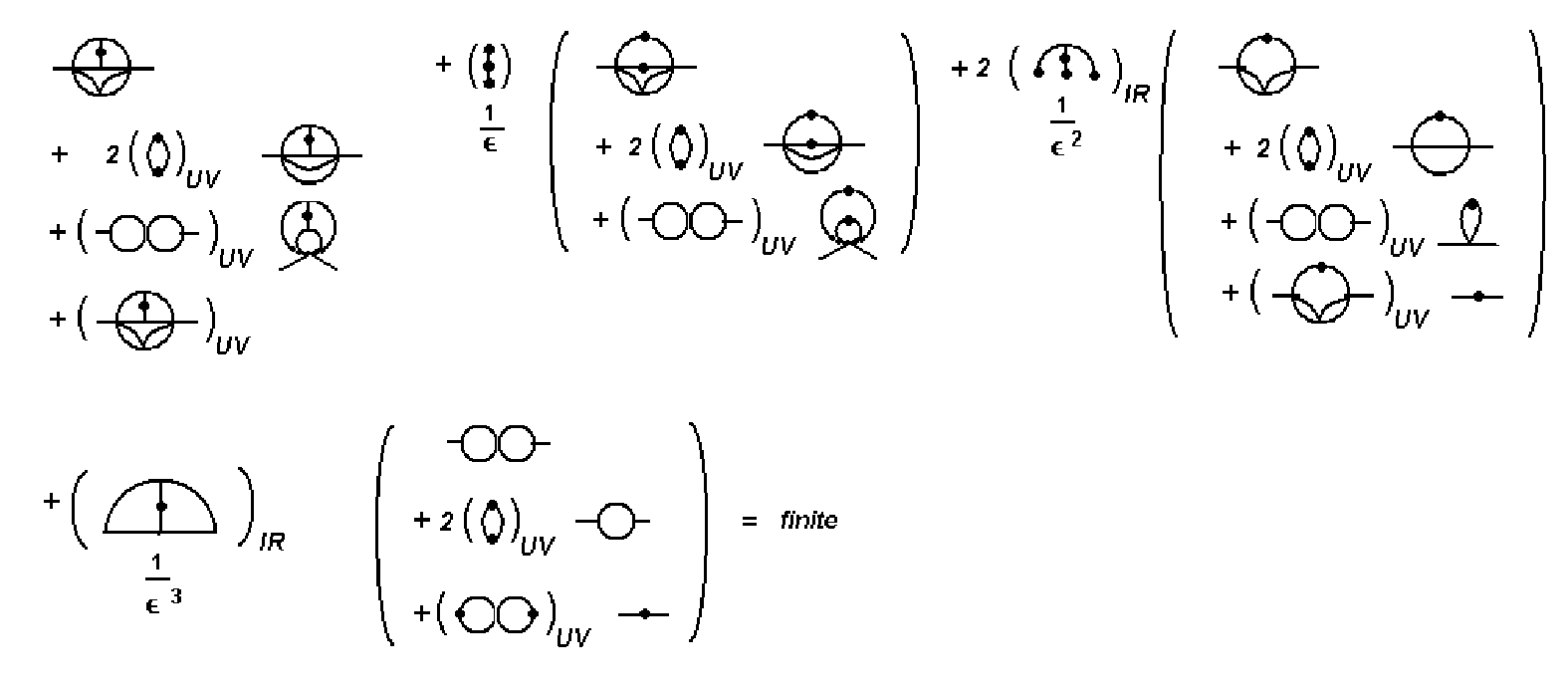}
\eqae Only the term at the bottom of the first column is needed
for the $\beta$ function.  Since we need all ${1 \over \e}$
poles, we must calculate each graph to some power in $\e$.  In
particular, since the IRD in the second column is proportional to
${1 \over \e}$ we need the graphs in this column to order $\e^0$,
but since the IRD in front of the third column is ${a \over \e^2}
+ {b \over \e}$, we need the graphs in this column to order $\e$,
and we need the terms proportional to $1, \e$ and $\e^2$ in the
graphs of the last column since the IRD in front of the last
column contains a leading term $1/\e^3$.  The three 1-loop,
2-loop and 3-loop tadpoles graphs in the first three columns in
the third row all vanish according to the rules of dimensional
regularization.

All these calculations were done using dimensional
regularization, not dimensional reduction (the latter is
inconsistent at higher loops~\cite{Avd}). For theories with
$\g_5$ and $\e_{\mu\nu\rho\sigma}$ the approach followed is to
first compute graphs without $\e$ tensors (substituting for
$\g_5$ the product $\e_{\mu\nu\rho\s} \g^\mu \g^\nu \g^\rho
\g^\s$), and only after the calculation is finished one contracts
with four-dimensional $\e$ tensors.  In fact, to compute $R_{UV}
F$ of a divergent graph with several $\e$ tensors, one can use
the fact that $R_{UV} F$ is $UV$ finite, and write the product of
two $\e$ tensors in terms of $D$-dimensional $\delta$ functions.
The error one makes is of order $\e$, so vanishes as $\e
\rightarrow 0$.  These $D$-dimensional Kronecker delta function
one can then insert inside the expression $R_{UV} F$ to obtain a
scalar.  If one has a single $\e_{\mu\nu\rho\s}$, one can
multiply by $k^\a_1 k^\beta_2 k^\g_3  k^\delta_4
\e_{\a\b\g\delta}$ and work out the product $\e_{\a\b\g\delta}
\e^{\mu\nu\rho\s}$ in terms of D-dimensional delta functions.
Effectively this means contracting the open indices in $R_{UV} F$
with momenta $k^\mu$ to obtain a Lorentz scalar. So, in the end
one never computes with open indices.  One obtains the correct
answer for the 3-loop chiral anomaly \cite{larin}.  This approach
works only for multiplicatively renormalizable quantities, and not
diagram-by-diagram.  The reason is that for multiplicatively
renormalizable models \eqa R_{UV} F = ZF \eqae Then the error in
using $D$-dimensional contractions is of order $D-4$.

Another example where the correct subtraction of IRD is crucial
for determining the UVD is the massless WZWN model in $D=2$
dimensions~\cite{dewitt}.  The simplest IR subtraction corresponds to
\eqa && R_{IR} \int {d^2 k \over k^2} = \int d^2 k \left( {1
\over k^2} + {\pi \over \e} \delta^2 (k) \right),\;\;\;\e=n-2 \nn
&& R_{IR}\;\;
\begin{fmffile}{fmfq30}\parbox{20mm}{
\begin{fmfgraph*}(12,12)
 \fmfleft{i}
 \fmfright{o}
 \fmf{vanilla}{i,v1}
 \fmf{vanilla}{v2,o}
 \fmf{vanilla,left,tension=.3}{v1,v2,v1}
\end{fmfgraph*}} \end{fmffile}  \hspace{-0.7cm}=
\begin{fmffile}{fmfq30}\parbox{20mm}{
\begin{fmfgraph*}(12,12)
 \fmfleft{i}
 \fmfright{o}
 \fmf{vanilla}{i,v1}
 \fmf{vanilla}{v2,o}
 \fmf{vanilla,left,tension=.3}{v1,v2,v1}
\end{fmfgraph*}} \end{fmffile}  \hspace{-0.7cm}+2
\left(\hspace{-0.3cm}
\begin{fmffile}{fmfpp7}\parbox{20mm}{
\begin{fmfgraph*}(10,6)
\fmftop{t} \fmfbottom{b} \fmfdot{t} \fmfdot{b}
 \fmf{vanilla}{t,b}
\end{fmfgraph*}}
\end{fmffile} \hspace{-1.3cm}\right)_{IR}
\begin{fmffile}{fmfpr1}
\parbox{20mm}{
\begin{fmfgraph*}(12,8)
  \fmfleft{i} \fmfright{o}\fmfdot{v}\fmfdot{v1} \fmf{vanilla}{i,v,v1,o}
  \end{fmfgraph*}}
\end{fmffile}\hspace{-0.7cm}
=\;\mbox{IR-finite} \eqae

A more complicated example is \setlength{\columnsep}{-200pt}
\begin{multicols}{2}
\includegraphics
[draft=false, height=1.2in,width=1.8in,keepaspectratio] {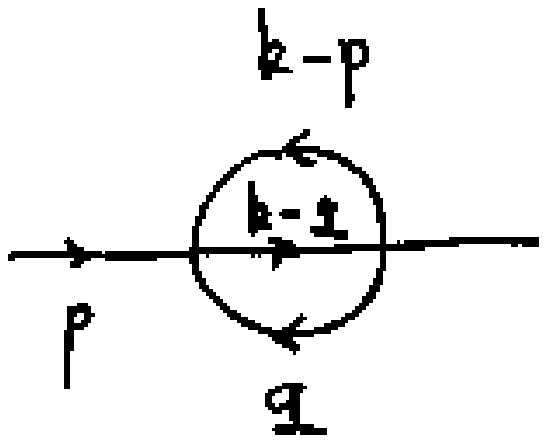}
\\

\begin{equation}
= \int {q_\rho (k - 2 q)_\s \over (k-q)^2 q^2 (k-p)^2} d^2 q d^2 p
\end{equation}
\end{multicols}
We write the numerator in terms of the momenta which appear in the
propagators, $q_\rho (k-2q)_\s = q_\rho (k-q)_\s -q_\rho q_\s$,
and obtain then graphically
\begin{equation}
\includegraphics
[draft=false, height=.7in,width=3.5in,keepaspectratio] {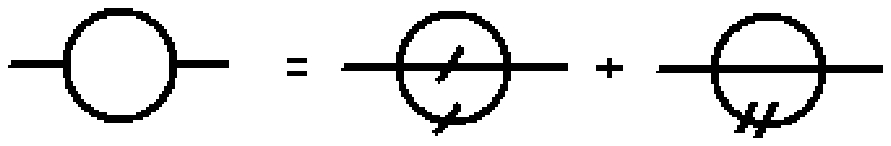}
\end{equation}
The slashes denote momenta, and reduce the IRD.  We obtain then
\begin{equation}
\includegraphics
[draft=false, height=1.2in,width=5.5in,keepaspectratio] {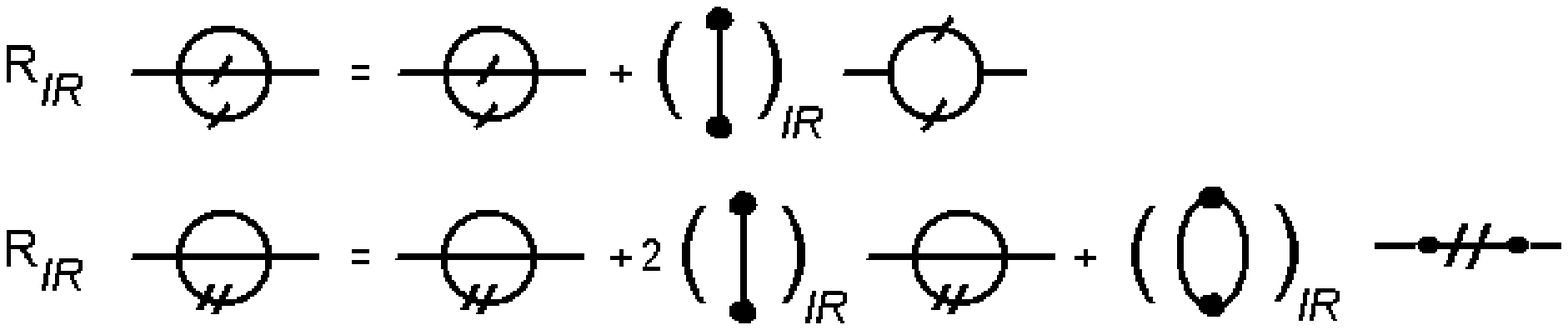}
\end{equation}
Having subtracted the IRD, one may then proceed to compute the
UVD. These are supposed to cancel in the $D=2$ WZWN model at $L
\geq 2$ loops, but one clearly needs to be careful with first
subtracting the correct amount of IRD.

If there are no momenta in the numerator, one obtains
\begin{equation}
\includegraphics
[draft=false, height=0.65in,width=5.5in,keepaspectratio] {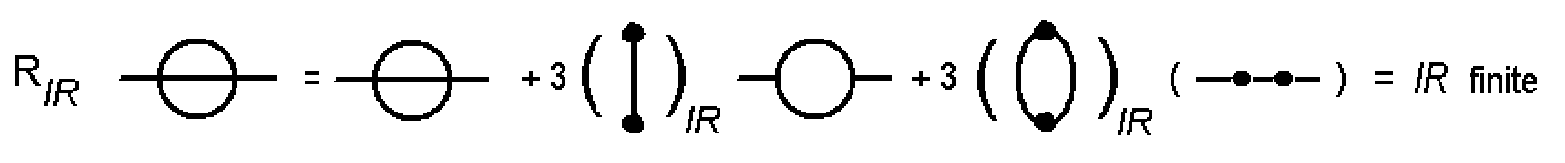}
\end{equation}
If one of the propagators is massive, we obtain (denoting the massive
propagator by a solid line)
\begin{equation}
\includegraphics
[draft=false, height=0.7in,width=5.5in,keepaspectratio] {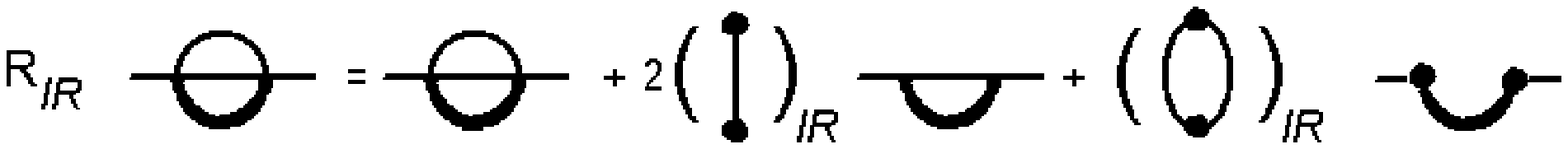}
\end{equation}

We close with two further examples. First consider in $D=4$ the
following 2-loop graph in massless $\lambda\varphi^4$
\begin{equation}
\includegraphics
[draft=false, height=.75in,width=1in,keepaspectratio] {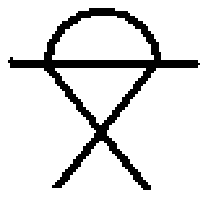}
\end{equation}
Suppose we want to compute the 2-loop contribution to the Z
factor of the $\varphi^4$ vertex. We first nullify two external
momenta because this simplifies the calculation
\begin{equation}
\includegraphics
[draft=false, height=0.8in,width=1.5in,keepaspectratio] {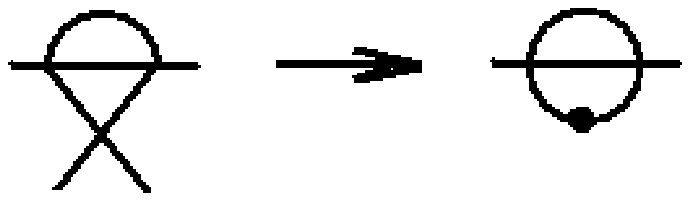}
\end{equation}
Next we subtract the IRD
\begin{equation}
\includegraphics
[draft=false, height=1in,width=5.5in,keepaspectratio] {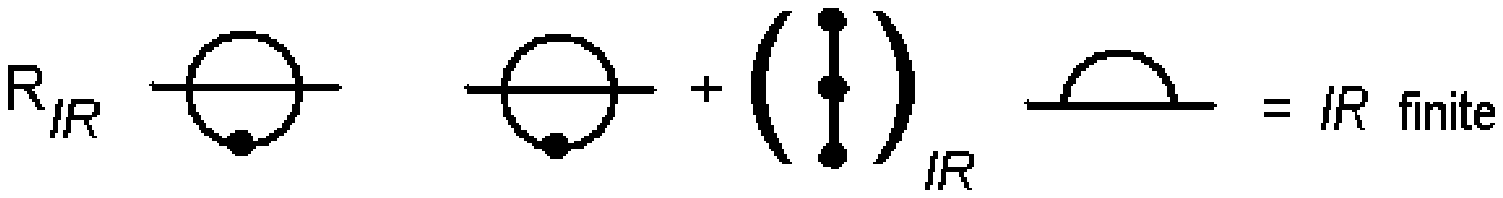}
\end{equation}
Finally we subtract the UVD
\begin{equation}
\includegraphics
[draft=false, height=1.5in,width=5.5in,keepaspectratio] {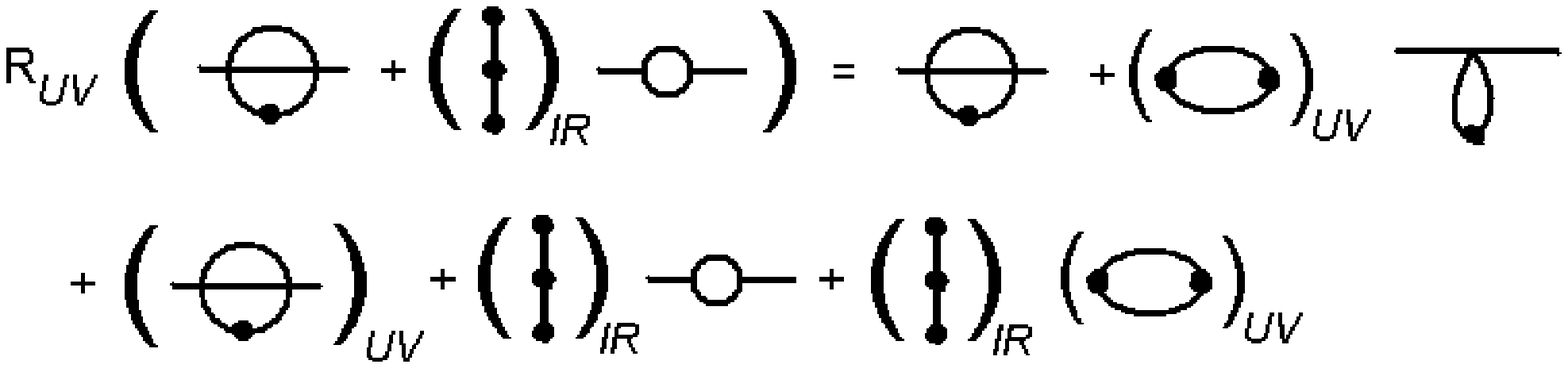}
\end{equation}
The tadpole graph vanishes, and the contribution to the Z factor
follows then from
\begin{equation}
\includegraphics
[draft=false, height=0.8in,width=5.5in,keepaspectratio] {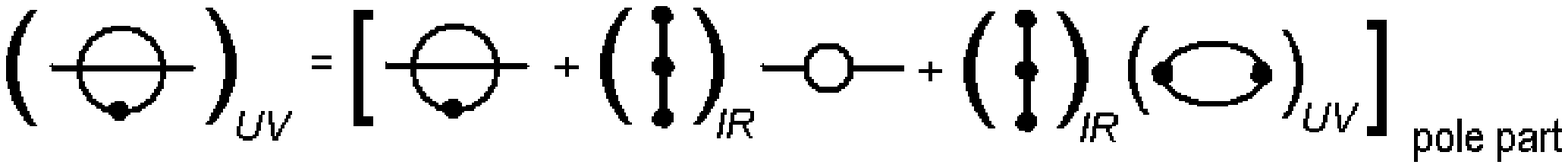}
\end{equation}

 For fun we give one last example \cite{vlad}.  We consider
the following 3-loop graph in massless $\l \varphi^4$ in $D=4$
\begin{equation}
\includegraphics
[draft=false, height=1.5in,width=5.5in,keepaspectratio] {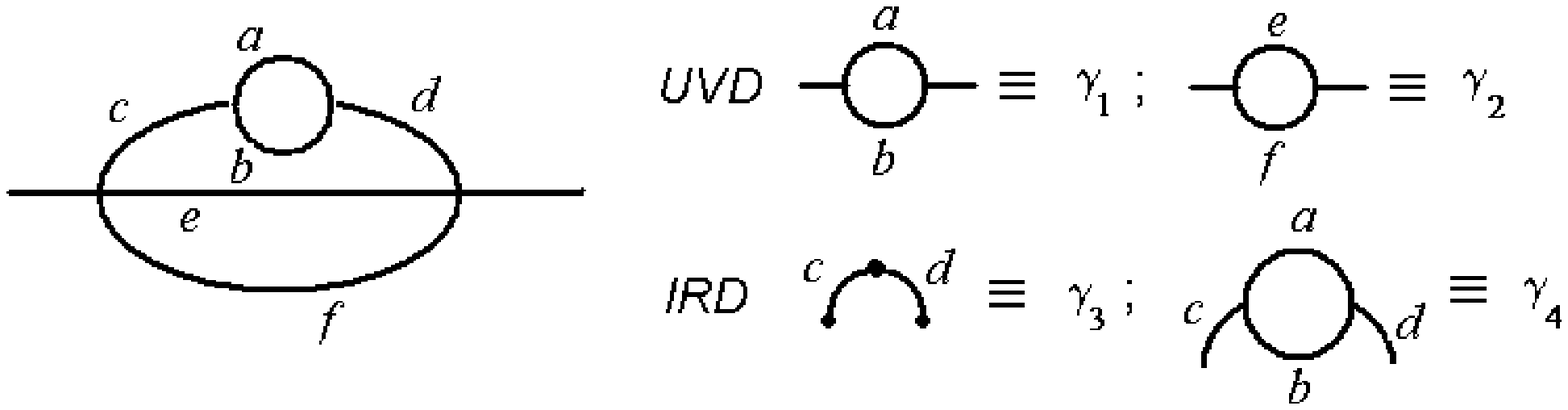}
\end{equation}
The infrared subtraction yields
\begin{equation}
\includegraphics
[draft=false, height=1in,width=5.5in,keepaspectratio] {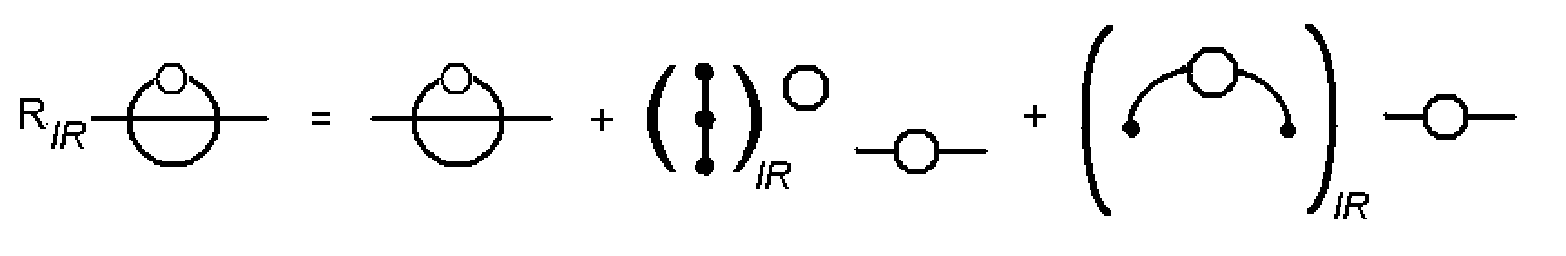}
\end{equation}
 Note the appearance of a
disconnected graph. Next ultraviolet subtraction yields
\begin{equation}
\includegraphics
[draft=false, height=4in,width=5.5in,keepaspectratio] {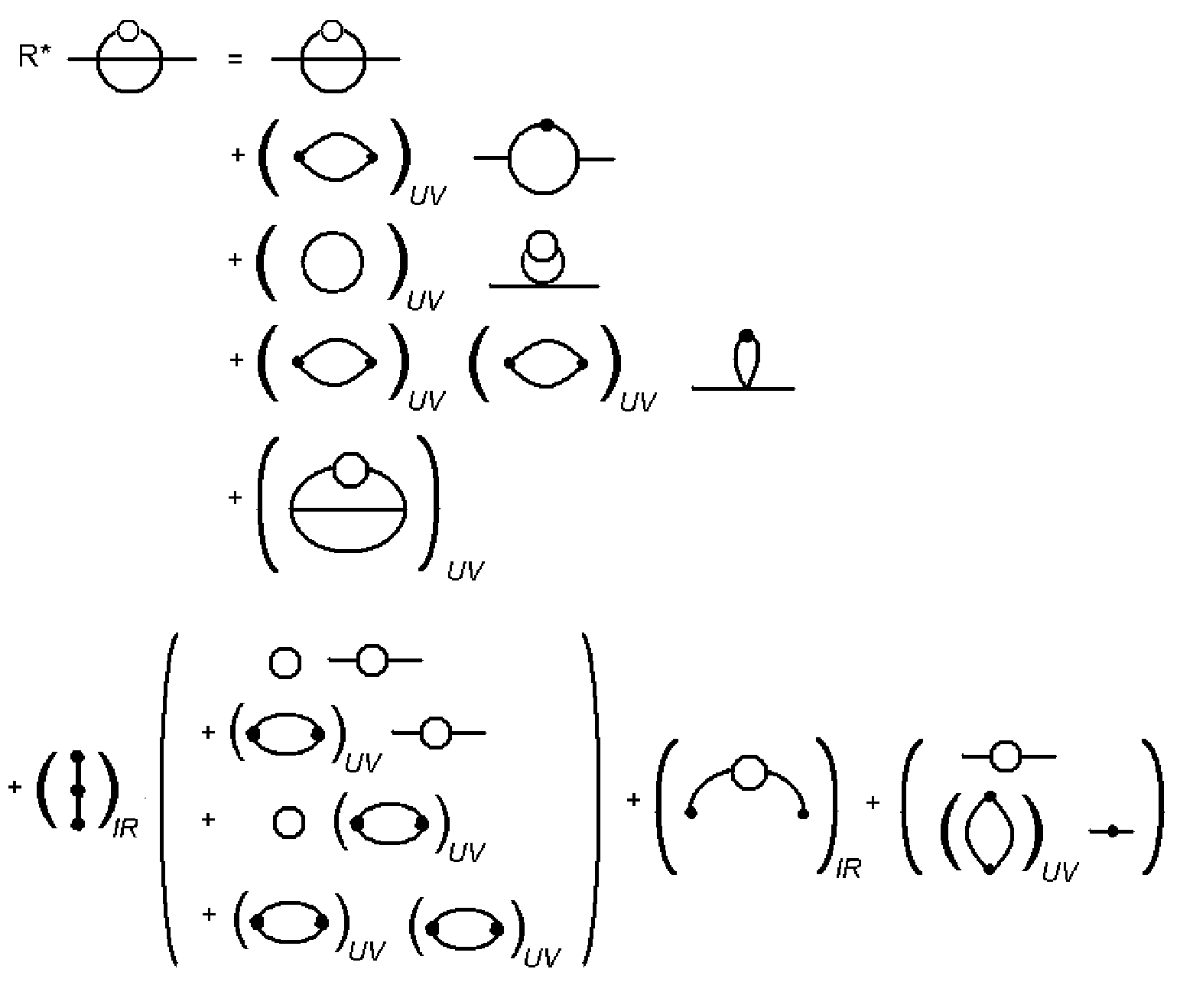}
\end{equation}
The UVD are $UVD  (\g_1) = UVD ( \g_2) = {-1 \over (4 \pi )^2
\e}$ with $\epsilon = n-4$, while the IRD are IRD $( \g_3) = {2
\over (4 \pi )^2 \e}$ and $IRD ( \g_4) $ is of the order $\frac{1}{\epsilon^2}$.  The overall UVD of
this graph is $P_G = {-(1- \e - \e^2) \over 3 \e^3 (4 \pi)^6}$
(where $G$ denotes the original graph), and $R^\ast G$ is UV and
IR finite.

We end with final conclusions and comments

\noindent (i) Graphs in which all masses have been set to zero
allow one to compute $\beta$ functions in a much simpler way than
keeping masses, but one introduces spurious IRD which one should
subtract. We have explained the rules for subtracting IRD from
Feynman graphs and for determining the final UVD which are
relevant for $\beta$ functions.\\
(ii) All UV counter terms and IR counter terms are polynomials in
${1 \over \e}$.  Tadpoles can be set to zero; even though they may
contain UVD; in the total result for the UVD as given by the
$R^\ast$
scheme no UVD are lost. \\
(iii)  One can write the counter terms as Feynman graphs with some
propagators replaced by $\delta^D (k)$ or $\square_k \delta^D
(k)$. For example, in $D=2$ one has $R_{IR} \int {d^2 k \over
k^2} = \int d^2 k ( {1 \over k^2} + {\pi \over \e} \delta^2
(k))$.  In this sense the IR counter terms are local in $p$
space.  In $x$-space they would be nonlocal, because infrared
divergences deal with the large $x$ behaviour.\\
(iv) To remove UVD one shrinks subgraphs, but to remove IRD one
deletes subgraphs. The final UVD is the one one needs for $\beta$
functions.  For lower loops $R_{UV} R_{IR}$ is equal to $R_{IR}
R_{UV}$, but for higher loops the order matters, and
the correct order is $R^\ast = R_{IR} R_{UV}$. The order only matters if a graph contains a factor
$\square_p \delta (p)$.\\
(v) in the original BPHZ approach~\cite{bog,zzub}, one starts
with $R_{UV} F = (1-t^F) \Pi_{H \e \Phi} (1- t^H)F$ where $H$ are
all proper subgraphs of the Feynman diagram $F$ which are
superficially divergent (power-counting divergent), while $t^F$
yields the overall divergence after all subdivergences have been
subtracted.  Furthermore, if $H \supset H'$ one should write $1-
t^H$ to the left of $(1-t^{H'})$, but if $H$ and $H'$ are
disjoint or overlapping, it does not matter in which order they
appear.  There exist refinements which show that one only needs
subsets of subgraphs which are nonoverlapping (``forests").  One
can then prove the forest formula $R_{UV} F = (1-t^F) \sum_i
\Pi_{H \e \phi_i} (-t^H) F$ where the $\phi_i$ are a forest which
includes the empty set.  In all examples above we have evaluated
this forest formula both for UVD and IRD.

\end{document}